\documentclass[a4paper,11pt]{article}
\usepackage{geometry}
\usepackage{a4wide}
\usepackage{graphicx}

\usepackage{amsmath}
\usepackage{amssymb}
\usepackage{soul}

\usepackage{cite}
\usepackage{float}
\usepackage{multirow,tabularx}
\usepackage{appendix}

\usepackage{tikz}
\usepackage[inline]{enumitem}
\usetikzlibrary{shapes}
\usepackage{verbatim}
\usepackage{hyperref}
\usepackage{graphicx,amsmath,amsfonts,amssymb,amsthm,euscript,braket,xcolor}
\newcommand{\be}{\begin{equation}}
\newcommand{\ee}{\end{equation}}

\newcommand{\Rmnum}[1]{\expandafter\@slowromancap\romannumeral #1@}
\newcommand{\bea}{\begin{eqnarray}}
\newcommand{\eea}{\end{eqnarray}}

\numberwithin{equation}{section}

\usepackage{subfigure}

\begin{document}
\title{\bf
A Thermodynamic Study of $\textbf{(2+1)}$-Dimensional Analytic Charged Hairy Black Holes with Born-Infeld Electrodynamics}

\author{\textbf{Shravani Sardeshpande}\thanks{shravanisardeshpande@gmail.com (422ph2080@nitrkl.ac.in)},~~\textbf{Ayan Daripa}\thanks{ayandaripaadra@gmail.com (419ma5024@nitrkl.ac.in)}
 \\\\
 \textit{{\small Department of Physics and Astronomy, National Institute of Technology Rourkela, Rourkela - 769008, India}}
}
\date{}
\maketitle
\abstract{}
This work presents analytical black hole solutions for a coupled Einstein-Born-Infeld-Scalar gravity system in AdS spacetime with two different non-minimal coupling functions
$f(z)$. For both solutions, we establish the regularity of the scalar field and curvature scalars outside the horizon. For one of the considered coupling cases, thermodynamic analysis in the canonical ensemble reveals stability across all temperatures, while the other case exhibits the Hawking/Page phase transition between the stable large phase of the black hole and thermal-AdS. We investigate the effect of the scalar hair parameter and black hole charge on the phase transition temperature and observe that the critical values of the scalar hair and the charge parameters constrain the feasibility of Hawking/Page phase transition.

\section{\label{sec1}Introduction}
The general theory of relativity has equipped us with a set of differential equations to describe the evolution of spacetime, and black holes, as solutions to those differential equations, have captured the attention of physicists for over a century. The discovery of black holes was a significant milestone in our understanding of the universe. However, the mysteries that these objects posed were a challenge to our understanding of physics. The laws of classical physics could no longer explain the behaviour of matter under such extreme conditions, which led people to look for new ways to understand black holes. The thermodynamics of black holes emerged as a fascinating subject that straddles both the classical and quantum aspects of gravity. The second law of thermodynamics necessitates that black holes have entropy\cite{Bekenstein:1973ur}. The study of black hole thermodynamics led to the realization that black holes behave like thermodynamic systems. They have temperature, emit radiation, and obey laws that are strikingly similar to the laws of thermodynamics \cite{Hawking:1975vcx, Gibbons:1976ue, Bekenstein:1973ur}. This realization opened up a new avenue to understand black holes and their behaviour.

Black hole thermodynamics expands into diverse spaces beyond flat spacetime, encompassing anti-de Sitter (AdS) and de sitter (dS) spaces. Within AdS spacetime, black holes exhibit notable thermodynamic stability, differing from asymptotically flat space. Additionally, they showcase intricate phase behaviours, including transitions like Hawking/Page phenomena (where transition occurs from stable large black hole to thermal-AdS) or van der Waals transitions (where transition occurs from small black hole to large black hole) \cite{Hawking:1982dh, Chamblin:1999tk, Chamblin:1999hg, Cvetic:1999ne, Dey:2015ytd, Mahapatra:2016dae}.

There has been growing interest in investigating simpler, lower-dimensional gravitational systems in recent years. The $(2+1)$-dimensional BTZ solutions studied extensively over the past three decades are a useful model for understanding black hole physics. It is described by topological field theory, which has a holographic correspondence with a two-dimensional CFT on its boundary \cite{Banados:1992wn, Banados:1992gq, Brown:1986nw}. This makes BTZ black holes ideal for testing gauge/gravity duality principles \cite{Maldacena:1997re}. Their entropy can be computed using conformal boundary conserved charges and symmetric algebra, offering insights into quantum gravity \cite{Strominger:1997eq}. Additionally, the  $(2+1)$-dimensional Chern–Simons gravity construction provides a grasp of the relation of gravity to gauge field theories \cite{Achucarro:1986uwr, Witten:1988hc}. BTZ black holes exhibit properties, such as the lack of curvature singularity and local equivalence to pure \(AdS_{3}\), distinct from their higher dimensional counterparts, making them more interesting to study. Also, thermodynamic and holographic interpretations make lower-dimensional models crucial in gravitational theories \cite{Carlip:1995qv}.

In Maxwell's electrodynamics, the field due to a point charge is given by Coulomb's law, which implies that the electric field tends to be infinite at the position of the particle. This leads to an infinite Lagrangian and self-energy for the point particle. In the 1930s, Max Born and Leopold Infeld introduced the Born-Infeld model \cite{Born:1934gh} to address the divergence of the self-energy of an electron in classical electrodynamics. To eliminate this infinity, an upper bound is imposed on the electric field, thus limiting the self-energy of the point charge \cite{Alam:2021ovb}. This was a significant departure from the traditional understanding and brought a new perspective to the field of electrodynamics. The Born-Infeld Lagrangian is given by
\begin{equation}
    \mathcal{L} = \frac{1}{2\alpha}\Big(1-\sqrt{1+\alpha \mathcal{F}^{2}}\Big),
\end{equation}
where $\mathcal{F}^{2}$ = $F_{\mu\nu}F^{\mu\nu}$, $F_{\mu\nu}$ is the electromagnetic field tensor, and $\alpha$ is the non-linearity parameter. The maximal possible value of the electric field in this theory is $1/2 \alpha$, and the self-energy of point charges is finite. The theory reduces to Maxwell electrodynamics under the limit $\alpha \rightarrow 0$. The nonlinear nature of this electrodynamics, which effectively eliminates the electric field singularity, led Deser and Gibbons to develop a Born-Infeld gravity theory using metric formulation \cite{Deser:1998rj}. Subsequently, Vollick brought in the Palatini approach \cite{Vollick:2003qp} to Born-Infeld gravity and explored its various facets. In \cite{Vollick:2005gc, Vollick:2006qd}, an interesting new method for coupling matter into this theory was introduced. It is worthwhile to note that the Born-Infeld action, when scalar fields are present, emerges as an effective action that controls the dynamics of vector fields on D-branes \cite{Asakawa:2012px, Callan:1997kz}. 

Considering a scalar field is intriguing due to its role in the low-energy limit of string theory, where it appears as a massless scalar field. This has sparked research into scalar gravity systems from various angles \cite{Mignemi:1991wa, Poletti:1994ww, Sheykhi:2007wg}. The scalar field significantly influences the causal structure and thermodynamic characteristics of charged black holes and alters spacetime geometry. With one or two Liouville-type scalar potentials, black hole solutions exhibit non-asymptotically flat or (anti)–de Sitter behaviours \cite{Mignemi:1991wa, Poletti:1994ww, Sheykhi:2007wg}. The coupling of a scalar field with other gauge fields can lead to significant changes in the resulting solutions \cite{Koikawa:1986qu, Gibbons:1987ps, Brill:1991qe}. Scalar fields also play a crucial role in forming black holes with unconventional asymptotes, such as charged Lifshitz black holes with an arbitrary dynamical exponent, requiring at least two scalar fields \cite{Tarrio:2011de}.

Building upon Martinez and Henneaux's seminal work, various solutions for hairy black holes with self-interacting real scalar fields in three spacetime dimensions have been explored \cite{Martinez:1996gn, Henneaux:2002wm}. This research has enhanced our understanding of the interaction between black hole geometries and scalar fields in $(2+1)$ dimensions, with implications for applications in holography. These scalar-gravity models are crucial for holography and offer valuable insights into two-dimensional condensed matter systems under strong couplings. The analytical solvability of $(2+1)$-dimensional gravity models containing scalar field has also piqued interest \cite{Chan:1994qa, Chan:1996rd, Ayon-Beato:2004nzi, Banados:2005hm, Correa:2010hf, Correa:2012rc, Xu:2013nia, Xu:2014uha, Cardenas:2014kaa, Tang:2019jkn, Dehghani:2017thu, Dehghani:2017zkm, Bueno:2021krl, Ahn:2015uza, Karakasis:2022fep, Zou:2014gla, Sadeghi:2013gmf, Zhao:2013isa, Bravo-Gaete:2014haa, Baake:2020tgk, Bravo-Gaete:2020ftn, Karakasis:2023ljt}. However, not all geometries constructed in $(2+1)$ dimensions exhibit desired physical characteristics. In most of the scenarios, the scalar field turns out to be logarithmic radial dependent, rendering the model unsuitable. Also, at the boundary, the geometry does not approach AdS spacetime  \cite{Dehghani:2017thu, Karakasis:2022fep}.

The scalar-gravity systems in three dimensions are also ideal for studying the no-hair theorem due to their ease of analysis \cite{Garcia-Diaz:2017cpv}. The no-hair theorem states that all stationary black hole solutions can be completely characterized by only three independent parameters: mass, electric charge, and angular momentum \cite{Ruffini:1971bza}. However, it is not a theorem in a strict mathematical sense, and many counterexamples exist, for instance \cite{Mavromatos:1999hp, Volkov:1989fi, Greene:1992fw, Kanti:1995vq, Luckock:1986tr, Droz:1991cx, Ovalle:2020kpd, Mahapatra:2022xea}. Therefore, it is interesting as well as desirable to look for analytic solutions for black holes using the Einstein-Born-Infeld-Scalar gravity system in $(2+1)$ dimensions. This has been done before in a few gravity systems; see, for instance, \cite{Dehghani:2022ntd, Mazharimousavi:2014vza}. However, some of these solutions contain drawbacks, as the scalar field depends logarithmically on the radial coordinate and, therefore, are undesirable, 

The presence of scalar hair can also significantly impact three-dimensional black holes' thermodynamic phase transition profile. In the case of four and higher dimensions, the multivaluedness of the entropy profile leads to different phases reflected in the temperature diagram. Like in the case of the Hawking/Page phase transition, the transition occurs between a stable black hole and thermal-AdS, but in the case of the van der Waals transitions or liquid/gas type phase transition, the transition happens between small and large black holes. On the contrary, the BTZ black hole shows a single entropy branch; hence, there are no phase transitions in either charged or uncharged cases. However, in recent studies, it was found that some three-dimensional hairy black hole solutions, constructed from the potential reconstruction technique, also undergo those phase transitions when nontrivial primary scalar hair is present \cite{Priyadarshinee:2023cmi, Daripa:2024ksg}.

When scalar and gauge fields are included in Einstein's gravity, it often leads to surprising and fascinating outcomes in the black hole geometry and their thermodynamics. Therefore, it is worthwhile to construct and investigate analytic solutions to the Einstein-Born-Infeld-Scalar gravity system in three dimensions with a regular scalar field profile and arbitrary coupling function and explore how the presence of a regular scalar field affects the geometrical and thermodynamical properties of black holes.

This work uses Born-Infeld electrodynamics to introduce analytical solutions for charged hairy black holes in $(2+1)$ dimensions. Our focus is primarily on the Einstein-Born-Infeld-Scalar gravity system, where there is a coupling function $f(z)$ between the scalar and gauge field. We have used the potential reconstruction method \cite{Dudal:2017max, Dudal:2021jav, Bohra:2020qom, Mahapatra:2018gig, Bohra:2019ebj, He:2013qq, Arefeva:2018hyo, Arefeva:2022avn, Arefeva:2020byn, Alanen:2009xs, Priyadarshinee:2021rch, Mahapatra:2020wym} in order to solve the Einstein-Born-Infeld-Scalar field equations. We have chosen a coupling function $f(z) = e^{-A(z)} \sqrt{1+\alpha ^2 q^4 z^4}$ and two different forms of the form factor are chosen to ensure a thorough analysis. We have considered (i) $A(z)$ = $ -\log(1+a^2 z^2)$, and (ii) $A(z)$ = $ -a^2 z^2$. Here, the parameter $a$ is associated with the strength of the scale hair. The considered form factors have been recently extensively studied for higher-dimensional hairy black holes in various contexts. Hence, we find it interesting to investigate the effects of the coupling functions on the geometry and the thermodynamics of $(2+1)$-dimensional black holes.

The structure of the paper is as follows: In Section 2, we discuss the $(2+1)$-dimensional Einstein-Born-Infeld-Scalar gravity model and present the exact solutions. Section 3 and 4 are devoted to examining the geometry and thermodynamics of the hairy black hole solutions with the form factors $A(z)= -\log(1+a^2 z^2)$ and $A(z)= -a^2 z^2$ and the coupling $f(z) = e^{-A(z)} \sqrt{1+\alpha ^2 q^4 z^4}$, respectively. Finally, in Section 5, we conclude and summarise our findings.

\section{\label{sec2}Black Hole Solution}
Let us start with the three-dimensional Einstein-Born-Infeld-Scalar action,
\begin{equation}
{S}=-\frac{1}{16\pi G_{3}}\int \,d^{3}x \sqrt{-g}\Big[R-\frac{1}{2}g^{\mu\nu}\partial_{\mu}\phi\partial_{\nu}\phi-V({\phi})+\frac{f(\phi)}{2\alpha}\Big(1-\sqrt{1+\alpha \mathcal{F}^{2}}\Big)\Big]\,.
\label{embiaction}
\end{equation}
Here, $R$ is the Ricci scalar,  $\phi$ is the scalar field,  $V(\phi)$ is the potential of the scalar field, the coupling function between the gauge and the scalar field is represented by $f(\phi)$, $\mathcal{F}^{2}$ is basically $F_{\mu\nu}F^{\mu\nu}$, where $F_{\mu\nu}$ denotes the Faraday tensor. We can represent $F_{\mu\nu}$ in terms of the four-potential $B_{\mu}$, i.e., $F_{\mu\nu}$= $\partial_{\mu}B_{\nu}-\partial_{\nu}B_{\mu}$. Here, $\alpha$ is a parameter that regulates the non-linearity of the Born-Infeld electrodynamics. Note that, as $\alpha \to 0$, the term $\mathcal{L}({\mathcal{F}})=\frac{1}{2\alpha}\Big(1-\sqrt{1+\alpha \mathcal{F}^{2}}\Big)$ reduces to the standard Maxwell electrodynamics which has the form, $\mathcal{L}({\mathcal{F}})=-\frac{1}{4}F_{\mu\nu}F^{\mu\nu}$. \newline

Eq.~\eqref{embiaction}, when varied with respect to the metric, gauge field, and scalar field, gives the Einstein, Born-Infeld, and scalar field equations, respectively. These are as follows:
\begin{equation}
\mathcal{R}_{\mu\nu}=-\frac{1}{2}g_{\mu\nu}V(z)+\frac{1}{2}\partial_\mu\phi \partial_\nu\phi-\frac{1}{4}g_{\mu\nu}\partial_\alpha\phi\partial^{\alpha}\phi +\frac{f(z)}{2}\Big[\frac{F_{\mu\alpha}F^{\alpha}_{\nu}}{\sqrt{1+\alpha\mathcal{F}^{2}}}+\frac{g_{\mu\nu}}{2\alpha}\left(1-\sqrt{1+\alpha\mathcal{F}^{2}}\right)\Big]\,,
\label{einsteineqn}
\end{equation}
\begin{equation}
\partial_\mu \biggl[\frac{\sqrt{-g}f(z)F^{\mu\nu}}{\sqrt{1+\alpha\mathcal{F}^{2}}} \biggr]=0\,,
\label{maxwelleqn}
\end{equation}
\begin{equation}
\partial_\mu[\sqrt{-g}\partial^{\mu}\phi]-\sqrt{-g}\Big[\frac{\partial V(z)}{\partial \phi}-\frac{1}{2\alpha}\frac{\partial f(z)}{\partial \phi}\left(1-\sqrt{1+\alpha\mathcal{F}^{2}}\right)\Big]=0\,.
\label{scalarfieldeqn}
\end{equation}
In order to obtain the $(2+1)-$dimensional hairy charged black hole solutions which are static and spherically ($S^1$) symmetric, we consider the following ans$\ddot{a}$tze  for the metric, scalar field solution, and gauge field solution:
\begin{eqnarray}
& & ds^2=\frac{e^{2A(z)} L^2}{z^2}\biggl[-g(z)dt^2 + \frac{  dz^2}{g(z)} + d\theta^2 \biggr]\,, \nonumber \\
& & \phi=\phi(z), \ \ B_{M}=B_{t}(z)\delta_{M}^{t} \, .
\label{metricansatz}
\end{eqnarray}
Here, $A(z)$ represents the scale factor, which plays an important role in describing the geometry and the thermodynamics of the black hole solution. $L$ denotes the length scale of AdS spacetime, which is set to unity for the ease of use, and $g(z)$ is the blackening function. The coordinate $z$ is the inverse of radial coordinate ($z=1/r$), which varies from $z = 0$ (which represents the AdS boundary) to $z= z_{h}$ (which represents the inverse of black hole horizon radius $z_{h}=1/r_{h}$) or to $z =\infty$ in the case of thermal-AdS solution (which is basically a horizon-less geometry). 

In the geometry specified by Eq.~\eqref{metricansatz}, there is only one non-zero component of Faraday's tensor, which is $F_{tz}$=$-B'_{t}(z)$. So we can write $\mathcal{F}^2=2F^2_{tz}g^{tt}g^{zz}=2B'^{2}_{t}(z)g^{tt}g^{zz}$. Now using Eq.~(\ref{maxwelleqn}), we get
\begin{equation}
F_{tz}=-B'_{t}(z)= -\frac{q e^{2 A(z)}}{z \sqrt{e^{2 A(z)} f(z)^2+2 \alpha  q^2 z^2}}\,,
\label{electricfield}
\end{equation}
where $q$ is an integration constant related to the charge of the black hole. Similarly, three Einstein equations of motion can be obtained by substituting Eq.~(\ref{metricansatz}) into Eq.~(\ref{einsteineqn}), which read as follows: 
\begin{equation}
\begin{aligned}
\hspace*{-0.5cm}tt\equiv & g'(z)\left(\frac{1}{2}A'(z)-\frac{1}{2z}\right)+g(z)\left(A''(z)+\frac{1}{z^2}+\frac{\phi'(z)^2}{4 }\right)+\frac{e^{2A(z)}V(z)}{2z^2}\\
& +\frac{f(z)}{2}\Big[\frac{z^{2}B_{t}'(z)^{2}e^{-2A(z)}}{\sqrt{1+\alpha\mathcal{F}^{2}}}-\frac{1}{2\alpha}\frac{e^{2A(z)}}{z^{2}}\left(1-\sqrt{1+\alpha \mathcal{F}^{2}}\right) \Big]=0\,,
\label{ttcomponent}
\end{aligned}
\end{equation}
\begin{equation}
\begin{aligned}
\hspace*{0.75cm} & zz\equiv g'(z)\left(\frac{A'(z)}{2} -\frac{1}{2 z}\right)+g(z)\left(A'(z)^2-\frac{2 A'(z)} {z}+\frac{1}{z^2}-\frac{\phi'(z)^2}{4} \right)\\
& +\frac{e^{2A(z)}V(z)}{2z^2}+\frac{f(z)}{2}\Big[\frac{z^{2}e^{-2A(z)}B_{t}'(z)^{2}}{\sqrt{1+\alpha\mathcal{F}^{2}}}-\frac{1}{2\alpha}\frac{e^{2A(z)}}{z^{2}}\left(1-\sqrt{1+\alpha \mathcal{F}^{2}}\right)\Big]=0\,,
\label{zzcomponent}
\end{aligned}
\end{equation}
\begin{eqnarray}
& & \theta\theta\equiv g''(z)+g'(z)\left(2A'(z)-\frac{2}{z}\right)+g(z)\left(2A''(z)+\frac{2}{z^2}+\frac{\phi'(z)^2}{2} \right)+\frac{e^{2A(z)}V(z)}{z^2}\nonumber\\
& &-\frac{f(z)}{2}\Big[\frac{1}{\alpha}\frac{e^{2A(z)}}{z^{2}}\left(1-\sqrt{1+\alpha \mathcal{F}^{2}}\right)\Big]=0 \,.
\label{thetathetacomponent}
\end{eqnarray}
In order to make them simpler to analyze, the above three Einstein equations can be further rearranged into the following equations,
\begin{equation}
g''(z)+g'(z)\left(A'(z)-\frac{1}{z}\right)-\frac{q^{2}e^{A(z)}}{\sqrt{f(z)^{2} e^{2A(z)}+2\alpha q^{2}z^{2}}}=0\,,
\label{g(z)eqn}
\end{equation}
\begin{equation}
A''(z)-A'(z)\left(A'(z)-\frac{2}{z}\right)+\frac{\phi '(z)^2}{2}=0\,,
\label{phi(z)eqn}
\end{equation}
\begin{eqnarray}
& &g''(z)+g'(z)\left(2A'(z)-\frac{2}{z}\right)+g(z)\left(2A''(z)+\frac{2}{z^2}+\frac{\phi'(z)^2}{2} \right)+\frac{e^{2A(z)}V(z)}{z^2}\nonumber\\
& &-\frac{f(z)}{2}\Big[\frac{1}{\alpha}\frac{e^{2A(z)}}{z^{2}}\left(1-\sqrt{1+\alpha \mathcal{F}^{2}}\right)\Big]=0 .
\label{v(phi)eqn}
\end{eqnarray}
Similarly, using Eq.~\eqref{scalarfieldeqn}, the scalar field equation of motion is given by,
\begin{equation}
  \phi''(z)+\phi'(z)\left(A'(z)+\frac{g'(z)}{g(z)}-\frac{1}{z}\right)-\frac{e^{2A(z)}}{z^{2}g(z)}\Big[\frac{\partial V(z)}{\partial \phi}-\frac{1}{2\alpha}\frac{\partial f(z)}{\partial \phi}\left(1-\sqrt{1+\alpha\mathcal{F}^{2}}\right)\Big]=0\,.
  \label{scalarEOM}
\end{equation}
Using Bianchi identity, we can check that, Eq.~(\ref{scalarEOM}) can be derived from equations~(\ref{g(z)eqn})-(\ref{v(phi)eqn}). Therefore, there are only four independent equations. Now, we impose the following boundary conditions to solve these equations:
\begin{eqnarray}
&& g(0)=1 \ \ \text{and} \ \ g(z_h)=0, \nonumber \\
%&& B_{t}(0)= \mu_e \ \ \text{and} \ \  B_{t}(z_h)=0, \nonumber \\
&& A(0) = 0 \,.
\label{boundarycdt2}
\end{eqnarray}
These conditions ensure that, at the boundary of the spacetime, i.e., at $z$ = 0, our chosen metric reduces to the usual AdS metric. We further demand that the blackening function $g(z)$ vanishes at the black hole horizon $z_h$. It is also crucial that the scalar field vanishes at the boundary, i.e., $\phi(0) = 0$, and remains real and finite in the spacetime in addition to these boundary criteria.

Using Eq.~(\ref{electricfield}) and applying the boundary conditions discussed above, we obtain the gauge field solution:
\begin{eqnarray}
B_{t}(z) = q \int_{z}^{z_h} \, d\xi~ \frac{e^{2 A(\xi)}}{\xi \sqrt{2 \alpha  q^{2}\xi^2+ e^{2 A(\xi)} f(\xi)^2}} \,.
\label{Btsol}
\end{eqnarray}
Similarly, using Eq.~(\ref{g(z)eqn}), the solution for $g(z)$ is found out to be
\begin{eqnarray}
& & g(z) =  C_1 + \int_0^z \, d\xi \ e^{-A(\xi)} \xi \biggl[ C_{2} + \mathcal{K}(\xi) \biggr] \,,
\label{gsol1}
\end{eqnarray}
with
\begin{eqnarray}
& & \mathcal{K}(\xi)= \int \, d\xi~ \frac{q^{2}e^{2A(\xi)}}{\xi \sqrt{e^{2A(\xi)}f^2(\xi)+2\alpha q^{2}\xi^{2}}}  \,,
\label{gsol2}
\end{eqnarray}
where $C_1$ and $C_2$ are the integration constants given by,
\begin{eqnarray}
C_1 = 1,  \ \ \ \ \ C_2 =- \frac{1+ \int_0^{z_h} \, d\xi~ e^{-A(\xi)} \xi \mathcal{K}(\xi) }{ \int_0^{z_h} \, d\xi~ e^{-A(\xi)} \xi}  \,.
\end{eqnarray}
Likewise, by solving Eq.~(\ref{phi(z)eqn}), the expression of scalar field $\phi$ is found to be
\begin{eqnarray}
\phi(z) = \int \, dz \ \sqrt{\frac{ - 4 A'(z)}{z}+ 2A'(z)^2- 2 A''(z)} + C_{3} \,,
\label{phisol}
\end{eqnarray}
where the integration constant $C_{3}$ can be obtained by recalling the condition that $\phi$ vanishes at the asymptotic boundary, i.e., $\phi |_{z=0}\rightarrow 0$. Finally, the expression for the potential $V(z)$ can be written from Eq.~(\ref{v(phi)eqn}) as follows:
\begin{eqnarray}
& & V(z) = g(z) \left(-2 z^2 e^{-2 A(z)} A''(z)-\frac{1}{2} z^2 e^{-2 A(z)} \phi'(z)^2-2 e^{-2 A(z)}\right)\nonumber\\
& & +\frac{f(z)}{2 \alpha }\left(1-\sqrt{1+\alpha\mathcal{F}^{2}}\right)+ 2z e^{-2 A(z)}\left(1 - z A'(z)\right) g'(z)- z^2 e^{-2 A(z)} g''(z) \,.
\label{Vsol}
\end{eqnarray}
Thus, we see that the $(2+1)$-dimensional Einstein-Born-Infeld-scalar gravity system can be
analytically solved and obtained solutions for the gauge field, blackening function, and scalar field in terms of the form factor $A(z)$ and the coupling function $f(z)$. Therefore, a family of closed-form analytic solutions can be easily obtained by selecting different forms of $A(z)$ and $f(z)$. Usually, in the gauge/gravity duality context, the forms of $A(z)$ and $f(z)$ are chosen to facilitate the study of the dual boundary field theory. For example, for the study of holographic QCD, the chosen forms of $A(z)$ and $f(z)$ must result in dual boundary field theory showing authentic QCD features, e.g., confinement/deconfinement phase transition \cite{Jain:2022hxl, Shukla:2023pbp, Jena:2022nzw}, confinement in the quark sector, linear Regge trajectory for the excited meson mass spectrum, etc.

However, without too much concern for the dual boundary field theory, we can adopt a more liberal and phenomenological approach and study different forms of $A(z)$ and $f(z)$ to discuss in detail the effects of scalar hair and to formulate a qualitative argument about the thermodynamics and stability of the hairy charged black holes in three dimensions with Born-Infeld electrodynamics. Here, we employ this strategy. In particular, we choose two different configurations of the form factor: (i) $A(z)$ = $-\log(1+a^2 z^2)$, and (ii) $A(z)$ = $-a^2 z^2$, and similarly choose a coupling function of type $f(z) = e^{-A(z)} \sqrt{1+\alpha ^2 q^4 z^4}$. These are chosen such as to make the computation of solutions of various geometric and thermodynamic quantities feasible. In particular, the various integrals appearing in Eqs. (\ref{Btsol}-\ref{Vsol}) can be straightforwardly evaluated for these choices of $f(z)$ and $A(z)$. Also, these forms of the scale factors $A(z)=-\log(1+a^2 z^2)$ have been studied extensively in the literature \cite{Priyadarshinee:2023cmi}. Particularly, the form of $A(z)=-a^2 z^2$ has also been widely studied in the holographic QCD literature; for instance, see \cite{Dudal:2017max, Bohra:2019ebj}. 
Here, the parameter $a$ regulates the strength of the scalar field. As a result, the scalar field back-reaction drops to zero as the parameter $a$ vanishes. Therefore, in the absence of scalar hair and gauge field, i.e., $a\rightarrow 0, q\rightarrow0$, our model simplifies to the standard BTZ black hole solution.

Our choice of aforementioned type of $f(z)$ and $A(z)$ can also be justified with following arguments:
\begin{itemize}
\item They guarantee that at the boundary $z\rightarrow 0$, the constructed hairy solutions asymptote to AdS. We have
\begin{eqnarray}
& & V(z)|_{z\rightarrow 0} = -\frac{2}{L^2} + \frac{m^2\phi^2}{2}+\dots \,, \nonumber\\
 & &  V(z)|_{z\rightarrow 0} =  2\Lambda + \frac{m^2\phi^2}{2}+\dots   \,,
\label{Vsolexp}
\end{eqnarray}
where the three-dimensional negative cosmological constant is $\Lambda=-1/L^2$, as is required. In the same way, the Ricci scalar $R$ asymptotically approaches $-6/L^2$. Together with the fact that $g(z)|_{z\rightarrow 0}=1$, this guarantees that the constructed solutions asymptote to AdS at the boundary. Furthermore, $m^{2}\geq-1$, the Breitenlohner-Freedman constraint \cite{Breitenlohner:1982jf} for stability in AdS space, is likewise satisfied by the mass of the scalar field $m^{2}=-1$. 

\item  Moreover, the obtained hairy solutions satisfy the Gubser criterion \cite{Gubser:2000nd} to have a well-defined dual boundary field theory.

\item It is also vital that the selected forms of $f(z)$ and $A(z)$ ensure that the null energy condition is always respected everywhere outside the horizon. This condition is expressed as
\begin{eqnarray}
T_{MN}\mathcal{N}^M \mathcal{N}^N \geqslant 0 \,,
\label{NEC}
\end{eqnarray}
where the null vector $\mathcal{N}^{M}$ satisfies the condition $g_{MN}\mathcal{N}^M \mathcal{N}^N=0$ and $T_{MN}$ is the energy-momentum tensor of the matter fields. The null vector $\mathcal{N}^{M}$ can be taken as
\begin{eqnarray}
\mathcal{N}^M= \frac{1}{\sqrt{g(z)}}\mathcal{N}^{t} + \cos{\beta}\sqrt{g(z)}\mathcal{N}^{z} + \sin{\beta} \mathcal{N}^{\theta}  \,,
\label{nullvector1}
\end{eqnarray}
for an arbitrary parameter $\beta$. Then, the null energy condition becomes,
\begin{eqnarray}
T_{MN}\mathcal{N}^M\mathcal{N}^N = \frac{z^{2}e^{-2A(z)}f(z) \sin^{2}\beta B'_{t}(z)^{2}}{2\sqrt{1+\alpha\mathcal{F}^{2}}}+\frac{1}{2}g(z)
   \cos ^2\beta \phi '(z)^2 \geqslant 0 \,.
\label{nullvector2}
\end{eqnarray}
which is always satisfied everywhere outside the horizon.
\end{itemize}

Let us now note the expressions of various thermodynamic observables of the obtained analytic black hole solutions. The expressions of the Hawking temperature and entropy are given by
\begin{equation}
\begin{aligned}
T=&\frac{z_{h}e^{-A(z_{h})}}{4\pi}\Big[-\mathcal{K}(z_{h})+ \frac{1+ \int_0^{z_h} \, d\xi~ e^{-A(\xi)} \xi \mathcal{K}(\xi) }{ \int_0^{z_h} \, d\xi~ e^{-A(\xi)} \xi}\Big],\\
&S_{BH}= \frac{\mathcal{A}}{4G_{3}}=\frac{L e^{A(z_{h})}}{4G_{3}z_{h}},
\label{tempandentropy}
\end{aligned}
\end{equation}
where $\mathcal{A}=2\pi/z_h$ is the area of the event horizon. 

The charge of the black hole can be obtained by measuring the flux of the electric field at the boundary, and it is given by
\begin{equation}
Q=\frac{1}{16\pi G_{3}}\int \frac{f(z)F_{\mu\nu}u^{\mu}n^{\nu}}{\sqrt{1+\alpha\mathcal{F}^{2}}} d\theta \,,
\label{electriccharge}
\end{equation}
where $u^{\mu}$ and $n^{\nu}$ are the unit space-like and time-like normals to the constant radial surface, respectively,
\begin{equation}
\begin{aligned}
u^{\mu}=&\frac{1}{\sqrt{-g_{tt}}}\delta^{\mu}_{t}=\frac{z}{Le^{A(z)}\sqrt{g(z)}}\delta^{\mu}_{t}\,,\\
n^{\nu}=&\frac{1}{\sqrt{g_{zz}}}\delta^{\nu}_{z}=\frac{z\sqrt{g(z)}}{Le^{A(z)}} \delta^{\nu}_{z}\,,
\label{normals}
\end{aligned}
\end{equation}
and $d\theta$ represents the integration across the one-dimensional boundary space. The actual value of the electric charge of the black hole, which is a function of the parameter $q$, is obtained using Eq.~\eqref{electricfield}
\begin{equation}
Q=\frac{q}{16\pi G_{3}}\,.
\label{simplifiedcharge}
\end{equation}

Also, another solution to the gravity equations exists that does not exhibit the horizon, called the thermal-AdS.\footnote{Here, this horizonless solution is referred to as thermal-AdS, for which the presence of non-vanishing energy-momentum tensor results in varying curvature throughout space-time.} The thermal-AdS solution corresponds to $g(z)=1$, and one may derive it from the black hole solution in the limit $z_{h}\to\infty$. Depending on the nature of $A(z)$, thermal-AdS can exhibit a non-trivial structure in bulk; however, due to the imposed boundary conditions (\ref{boundarycdt2}), and similar to the case of a black hole, it too always asymptotes to AdS at the boundary. In the further sections of this work, we focus on the possibility of phase transition between the phases of the black hole and thermal-AdS which is also known as the Hawking/Page phase transition.

\section{\label{sec:f=phi/2} Solution of the hairy black hole with form factor $A(z)=-\log(1+a^2 z^2)$}

In this section, we look into the geometry and thermodynamics of the black hole solution with the coupling and form factor $f(z)$ = $e^{-A(z)} \sqrt{1+\alpha ^2 q^4 z^4}$ and $A(z)=-\log(1+a^2 z^2)$, respectively. Using Eq.~(\ref{phisol}), we obtain the scalar field solution as follows:
\begin{equation}
\phi(z)=2 \sqrt{3} \sinh ^{-1}(a z) \,.
\label{phivalue1}
\end{equation}
Using Eq.~\eqref{electricfield}, we obtain
\begin{equation}
  F_{tz}= -\frac{q}{\left(a^2 z^2+1\right)^2 \left(\alpha  q^2 z^3+z\right)}.
\end{equation}
From there, the gauge field solution comes out to be
\begin{equation}
    \begin{aligned}
         B_t(z) &=\frac{q}{2}  \left(\frac{\alpha ^2 q^4 \log \left(\frac{\alpha  q^2 z^2+1}{\alpha  q^2 z_h^2+1}\right)}{\left(a^2-\alpha  q^2\right)^2}+\frac{a^4 \left(z-z_h\right) \left(z_h+z\right)}{\left(a^2 z^2+1\right) \left(a^2 z_h^2+1\right) \left(a^2-\alpha  q^2\right)}\right)\\
         &\hspace{0.3cm}+q \left(\log \left(z_h\right)-\log (z)+\frac{\left(a^4-2 a^2 \alpha  q^2\right) \log \left(\frac{a^2 z^2+1}{a^2 z_h^2+1}\right)}{2\left(a^2-\alpha  q^2\right)^2}\right) \ . 
        \label{gaugefield1}
    \end{aligned}
\end{equation}
When $a\to0$, i.e. the scalar field vanishes, and the gauge field reduces to 
\begin{equation}
   B_t(z)\big\rvert_{a \rightarrow0} = \frac{q}{2}  \left(\log \left(\frac{\alpha  q^2 z^2+1}{\alpha  q^2 z_h^2+1}\right)-2 \log \left(\frac{z}{z_h}\right)\right).
\end{equation}
Now, with the help of Eq.~\eqref{g(z)eqn}, the expression for the blackening function can be obtained as
\begin{equation}
    \begin{aligned}
         g(z) &=1+\frac{q^2 \left(a^2 z^2 \left(a^2-\alpha  q^2\right)+2 z^2 \left(a^2 z^2+2\right) \log (z) \left(a^2-\alpha  q^2\right)^2\right)}{8\left(a^2-\alpha  q^2\right)^2}\\
         &-\frac{q^2\left(\left(\alpha  q^2 z^2+1\right) \left(\alpha  q^2 \left(a^2 z^2+2\right)-a^2\right) \log \left(\alpha  q^2 z^2+1\right)\right)}{8 \left(a^2-\alpha  q^2\right)^2}\\
         &-\frac{z^2 \left(a^2 z^2+2\right) q^2  \log ({z_h})  }{4  }-\frac{z^2 \left(a^2 z^2+2\right)q^2 a^2  }{8  \left(a^2-\alpha  q^2\right) \left(a^2 {z_h}^2+2\right)}\\
         &+\frac{z^2 \left(a^2 z^2+2\right)q^2 \left(\alpha  q^2 {z_h}^2+1\right) \left(\alpha  q^2 \left(a^2 {z_h}^2+2\right)-a^2\right) \log \left(\alpha  q^2 {z_h}^2+1\right)}{8  {z_h}^2\left(a^2-\alpha  q^2\right)^2 \left(a^2 {z_h}^2+2\right)}\\
         &-\frac{{z}^2 \left(a^2 {z}^2+2\right)}{{z_h}^2 \left(a^2 {z_h}^2+2\right)}-\frac{q^2\left(a^2 z^2+1\right)^2 \left(a^2-2 \alpha  q^2\right) \log \left(a^2 z^2+1\right)}{8 \left(a^2-\alpha  q^2\right)^2}\\
         &+\frac{z^2 \left(a^2 z^2+2\right)q^2\left(a^2 {z_h}^2+1\right)^2 \left(a^2-2 \alpha  q^2\right)\log \left(a^2 {z_h}^2+1\right)}{8 {z_h}^2 \left(a^2-\alpha  q^2\right)^2 \left(a^2 {z_h}^2+2\right)}\,.
        \label{g(z)1}
    \end{aligned}
\end{equation}
Note that in the absence of scalar hair, the above equation simplifies to that of the standard charged non-hairy BTZ black hole coupled with Born-Infeld type of gauge field \cite{Myung:2008kd}, i.e.,
\begin{eqnarray}
    \begin{aligned}
    g(z)\big\rvert_{a \rightarrow0}&=\frac{\left(2 \alpha  q^2 z^2 \log \left(\frac{z}{z_h}\right)-\left(\alpha  q^2 z^2+1\right) \log \left(\alpha  q^2 z^2+1\right)\right)}{4 \alpha  }\\
    & + \frac{ \left(z_{h}^2-z^2\right)}{z_{h}^2}+\frac{z^2 \left(\alpha  q^2 {z_h}^2+1\right) \log \left(\alpha  q^2 {z_h}^2+1\right)}{4 \alpha  {z_h}^2}\,.
    \end{aligned}
\end{eqnarray}
%%%%%%%%%%%%%%%%%%%%%%%%%%%%%%

\begin{figure}[h!]
\begin{minipage}[b]{0.5\linewidth}
\centering
\includegraphics[width=2.8in,height=2.3in]{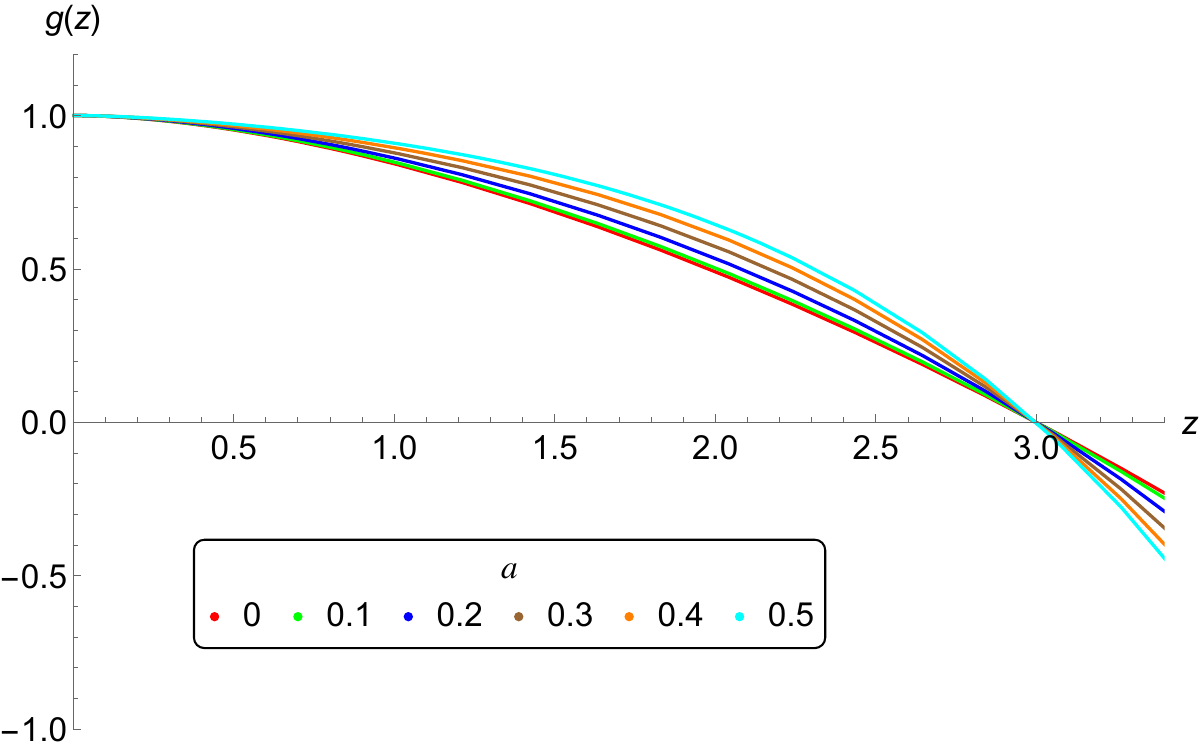}
\caption{ \small The radial profile of the blackening function for varying $a$ keeping $z_{h}=3.0$, $\alpha=0.5$, and $q=0.3$ fixed.}
\label{GvsZaF1qPt3alphaPt5}
\end{minipage}
\hspace{0.4cm}
\begin{minipage}[b]{0.5\linewidth}
\centering
\includegraphics[width=2.8in,height=2.3in]{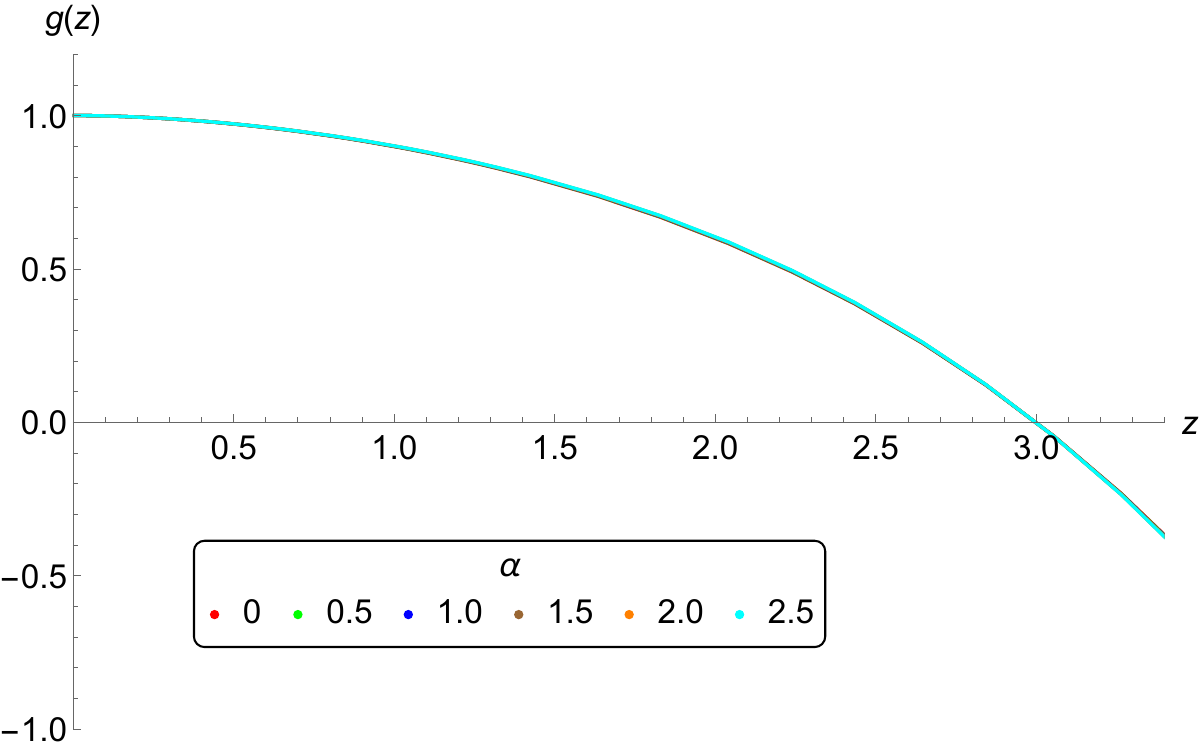}
\caption{\small The radial profile of the blackening function for varying $\alpha$ keeping $z_{h}=3.0$, $a=0.3$, and $q=0.2$ fixed.}
\label{GvsZalphaF1aPt3qPt2}
\end{minipage}
\end{figure}

\begin{figure}[h!]
\begin{minipage}[b]{0.5\linewidth}
\centering
\includegraphics[width=2.8in,height=2.3in]{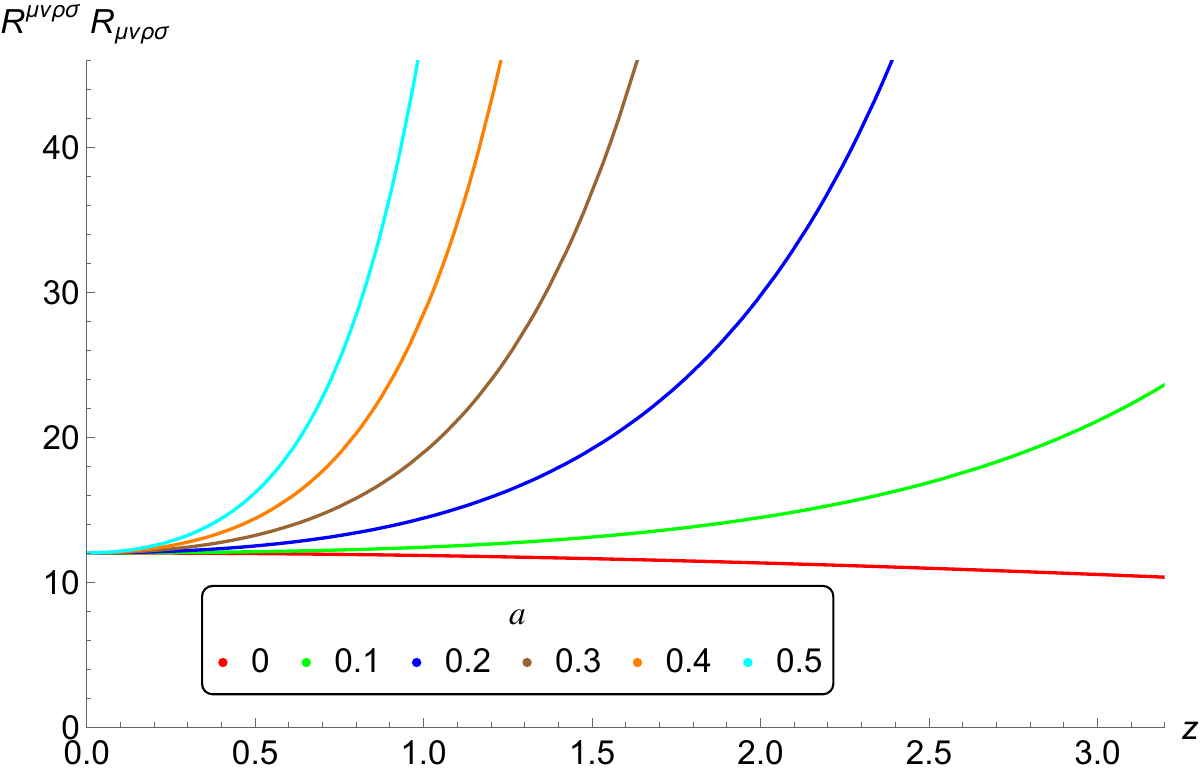}
\caption{ \small The nature of $R_{\mu\nu\rho\sigma}R^{\mu\nu\rho\sigma}$ for varying $a$ keeping $z_{h}=3.0$, $\alpha=0.5$, and $q=0.3$ fixed.  }
\label{KSvsZaF1qPt3alphaPt5}
\end{minipage}
\hspace{0.4cm}
\begin{minipage}[b]{0.5\linewidth}
\centering
\includegraphics[width=2.8in,height=2.3in]{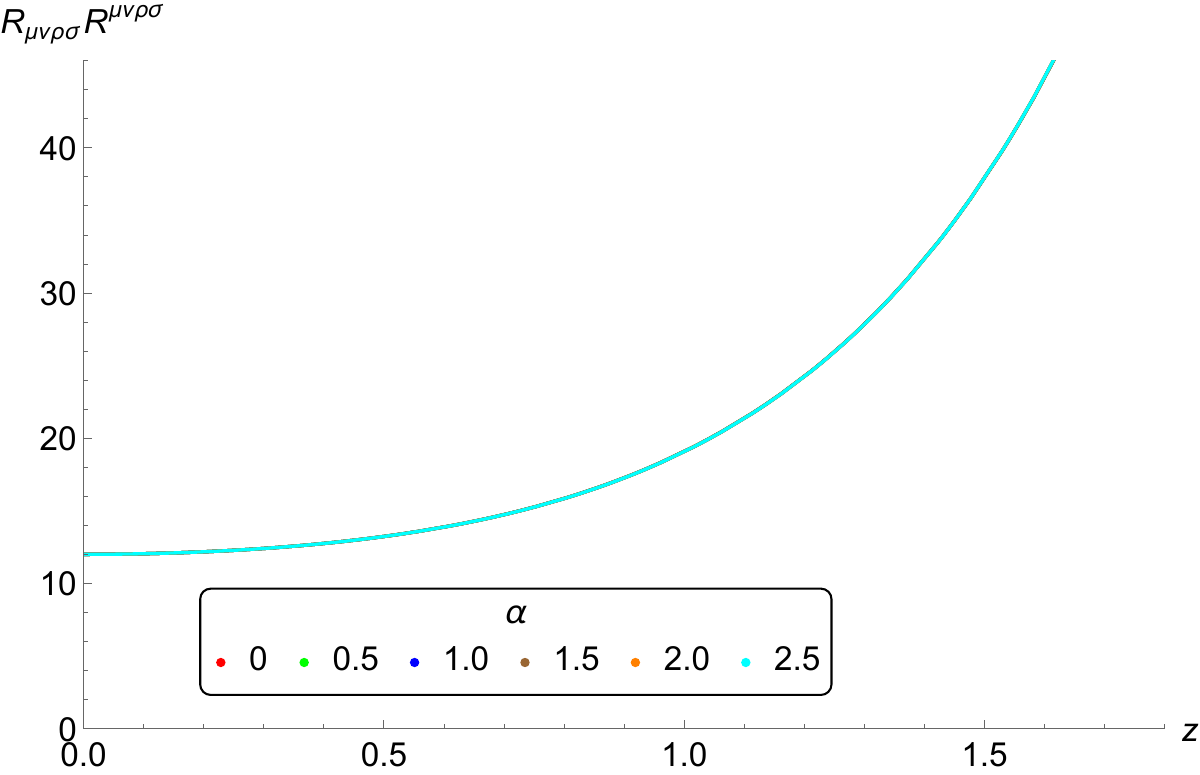}
\caption{\small The nature of $R_{\mu\nu\rho\sigma}R^{\mu\nu\rho\sigma}$ for varying $\alpha$ keeping $z_{h}=3.0$, $a=0.3$, and $q=0.2$ fixed.}
\label{KSvsZalphaF1aPt3qPt2}
\end{minipage}
\end{figure}
\begin{figure}[h!]

\subfigure[]{\includegraphics[width=0.5\textwidth]{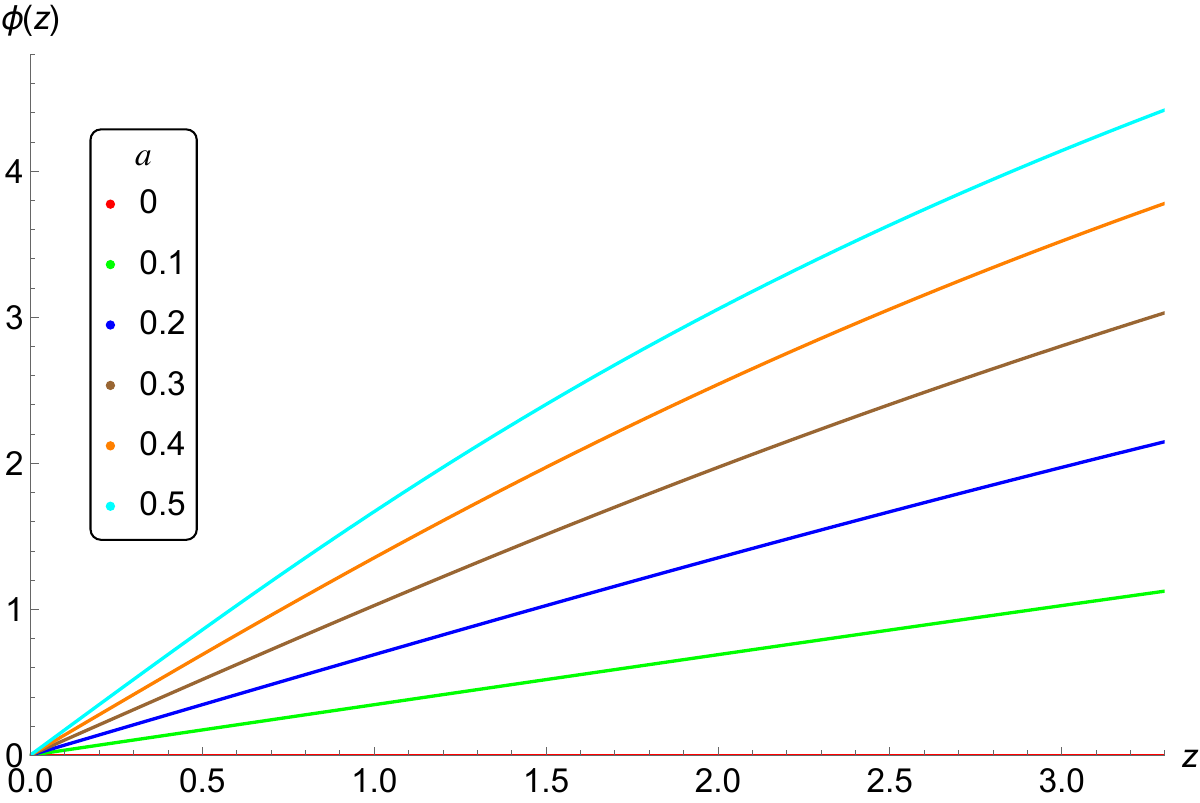}}
    \subfigure[]{\includegraphics[width=0.5\textwidth]{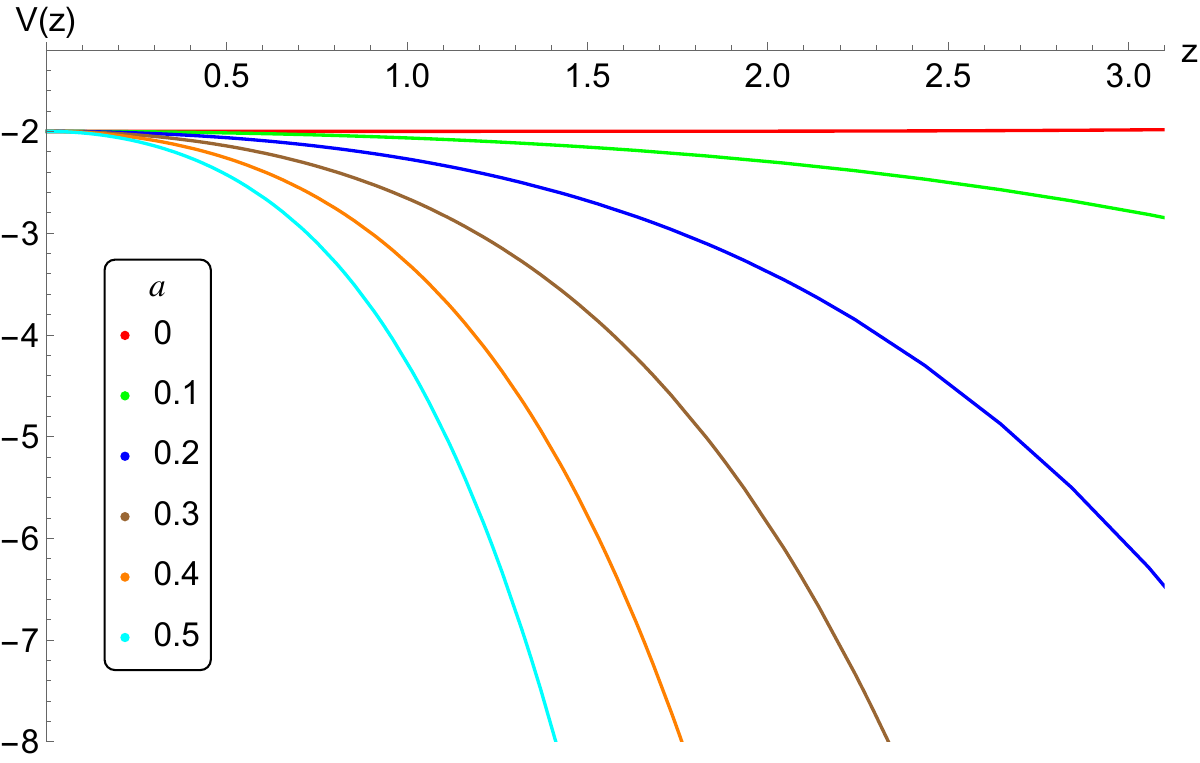}} 
    \caption{The radial nature of the scalar field and its potential for varying $a$ keeping $z_{h}=3.0$, $\alpha=0.5$, and $q=0.3$ fixed. }
    \label{phi_V_F1}
\end{figure}
%%%%%%%%%%%%%%%%%%%%%%%%%%%%%%

We now present our results for the geometrical features of this hairy black hole solution. Our results are presented in Figs. (\ref{GvsZaF1qPt3alphaPt5}-\ref{phi_V_F1}). Here, we have fixed the value of $z_h=3$ while presenting our results throughout this work. Fig.~\ref{GvsZaF1qPt3alphaPt5} shows the radial profile of the blackening function $g(z)$ for various values of the scalar hair parameter $a$ keeping $\alpha=0.5$ and $q=0.3$. Similarly, in Fig.~\ref{GvsZalphaF1aPt3qPt2}, we have plotted the nature of $g(z)$ as a function of $z$ for different values of the Born-Infeld non-linearity parameter $\alpha$ when $a=0.3$ and $q=0.2$.
Note that, at $z=z_h$, a change in the sign of $g(z)$ is observed, thereby confirming the presence of the horizon. This assertion holds validity irrespective of the value assigned to $a$. We also observe that there is no significant dependence of $g(z)$ on $\alpha$. The Figs.~\ref{KSvsZaF1qPt3alphaPt5} and \ref{KSvsZalphaF1aPt3qPt2} display the radial profiles of the Kretschmann scalar $R_{\mu\nu\rho\sigma}R^{\mu\nu\rho\sigma}$ for various values of $a$ and $\alpha$. We observe that the curvatures scalar, like the Kretschmann scalar and the Ricci scalar, remain finite throughout spacetime. The curvature singularity, concealed by the event horizon, exists only where the Kretschmann scalar diverges at $z=\infty$ or $r=0$. Therefore, we can conclude that, when compared with the non-hairy charged BTZ black hole, the addition of hairiness to the charged black hole does not affect the number of singularities present.

The scalar field profile is observed to be finite and real at all points at the horizon and beyond and goes to zero only at the asymptotic boundary. This indicates the existence of a stable hairy black hole solution exhibiting Born-Infeld electrodynamics in a three-dimensional space. 
Similarly, we can obtain the expression of the potential $V(z)$, the radial profile of which is displayed in Fig.~\ref{phi_V_F1}. The potential in the area outside the horizon is also regular and finite. The potential approaches $V(z=0)=-2/L^2$ at the boundary for all values of $a$, $\alpha$, and $q$.  From the plot, it is obvious that the potential is bounded from above. This satisfies the Gubser criterion to establish a well-defined boundary field theory \cite{Gubser:2000nd}. However, the Gubser criterion is not respected for larger values of $q \gtrsim 1.5$. So, in this work, we will restrict ourselves only to small values of the charge parameter to avoid violating the criterion.

Now, let us move on to the thermodynamic analysis of this black hole solution. Using Eq.~(\ref{tempandentropy}), the expression of the black hole temperature comes out to be:
\begin{equation}
    \begin{aligned}
        T=&\frac{a^2 z_h \left(\left(2 \alpha  q^4-a^2 q^2\right) \log \left(\frac{a^2 z_h^2+1}{\alpha  q^2 z_h^2+1}\right)+\left(a^2-\alpha  q^2\right) \left(8 a^2-(8 \alpha +1) q^2\right)\right)}{8 \pi  \left(a^2 z_h^2+2\right) \left(a^2-\alpha  q^2\right)^2}\\
        &+  \frac{1}{\pi  z_h \left(a^2 z_h^2+2\right)}+\frac{\left(2 \alpha  q^4-a^2 q^2\right) \log \left(\frac{a^2 z_h^2+1}{\alpha  q^2 z_h^2+1}\right)}{8 \pi  z_h \left(a^2 z_h^2+2\right) \left(a^2-\alpha  q^2\right)^2}.
    \end{aligned}
\end{equation}
%%%%%%%%%%%%%%%%%%%%%%%%%%%%%%
\begin{figure}[h!]
\begin{minipage}[b]{0.5\linewidth}
\centering
\includegraphics[width=2.8in,height=2.3in]{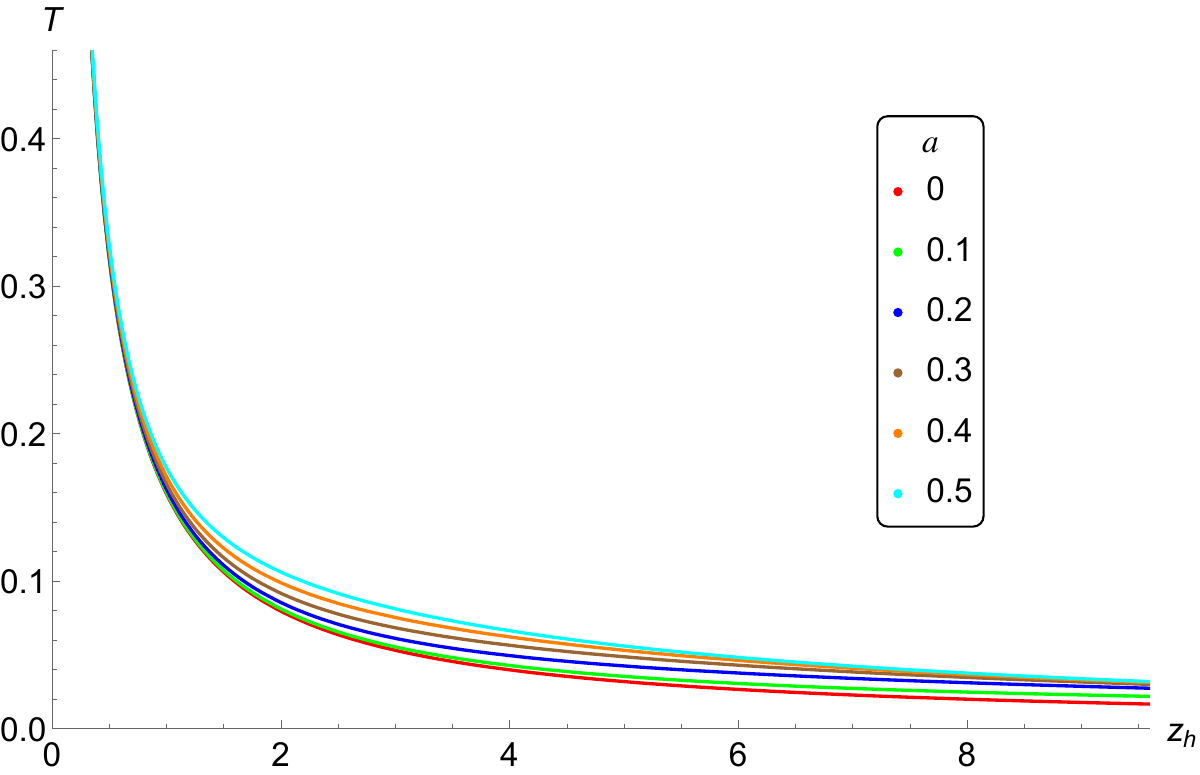}
\caption{ \small The nature of black hole temperature, $T$, plotted against the inverse horizon radius $z_h$ for varying $a$ keeping $q=0$.  }
\label{Tvszhcase1a}
\end{minipage}
\hspace{0.4cm}
\begin{minipage}[b]{0.5\linewidth}
\centering
\includegraphics[width=2.8in,height=2.3in]{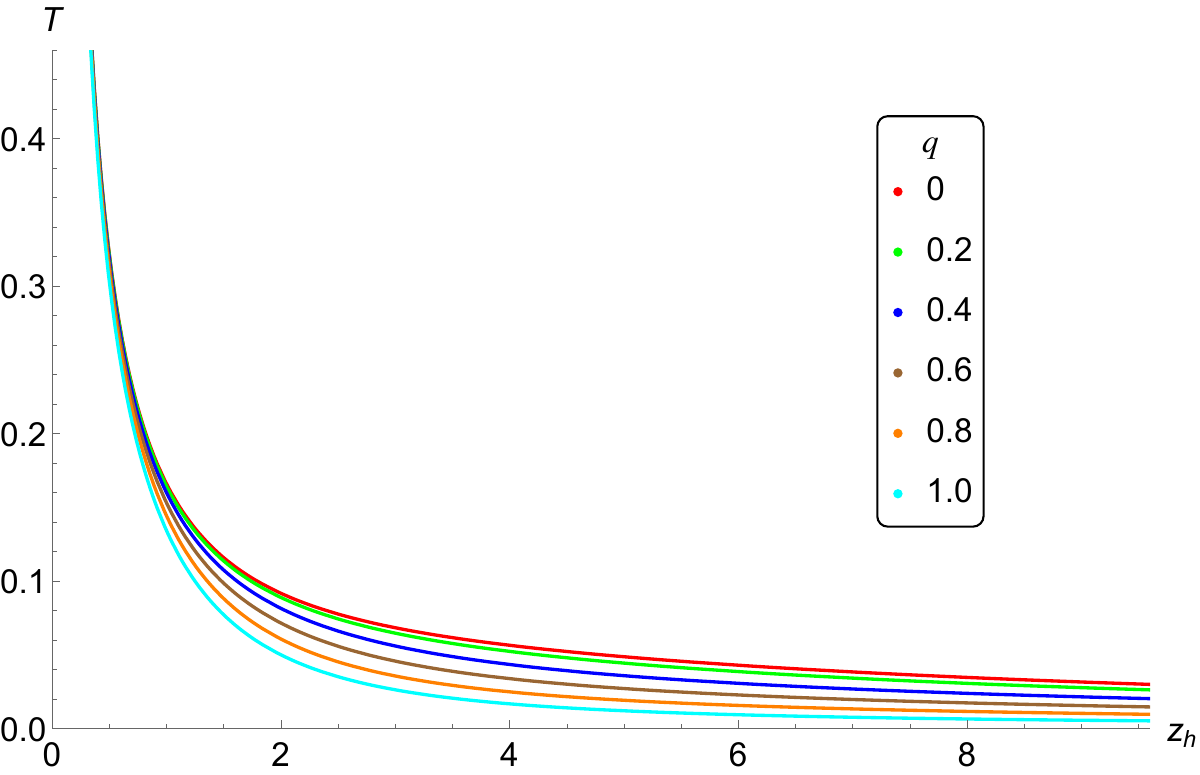}
\caption{ \small The nature of black hole temperature, $T$, plotted against the inverse horizon radius $z_h$ for varying $q$ keeping $a=0.3$ and $\alpha=0.5$ fixed.}
\label{tvszhcase1-q}
\end{minipage}
\end{figure}
%%%%%%%%%%%%%%%%%%%%%%%%%%%%%
In the absence of a scalar field, the above temperature expression smoothly reduces to that of Born-Infeld BTZ solution, i.e., 
\begin{equation}
T\big\rvert_{a \rightarrow0} = \frac{1}{2 \pi  z_h}-\frac{\log \left(\alpha  q^2 z_h^2+1\right)}{8 \pi  \alpha  z_h}.
\label{tempcase1a0}
\end{equation}
In Fig.~\ref{Tvszhcase1a}, the Hawking temperature of the black hole is plotted against the inverse horizon radius $z_h=1/r_h$ for small sequential values of $a$ with $q=0$ fixed. As a result, for uncharged hairy black holes, the Born-Infeld parameter $\alpha$ does not play any role. We observe that, for all values of $a$, only one black hole phase exists, which is thermodynamically stable at all temperatures. Thus, no phase transition is observed for this black hole solution. Also, note that the temperature of this black hole phase decreases with an increase in inverse horizon radius $z_h$, resulting in a positive value for the specific heat.  Consequently, in this case, the thermodynamics of the hairy solution resembles the thermodynamic properties of the uncharged BTZ black hole. 

In Fig.~\ref{tvszhcase1-q}, the temperature profile is plotted against the $z_h$ for different values of $q$ keeping $a=0.3$ and $\alpha=0.5$ fixed. In this case as well, we observe that only one thermodynamically stable phase exists at all temperatures for all values of $q$.

\begin{figure}[h!]
\begin{minipage}[b]{0.5\linewidth}
\centering
\includegraphics[width=2.8in,height=2.3in]{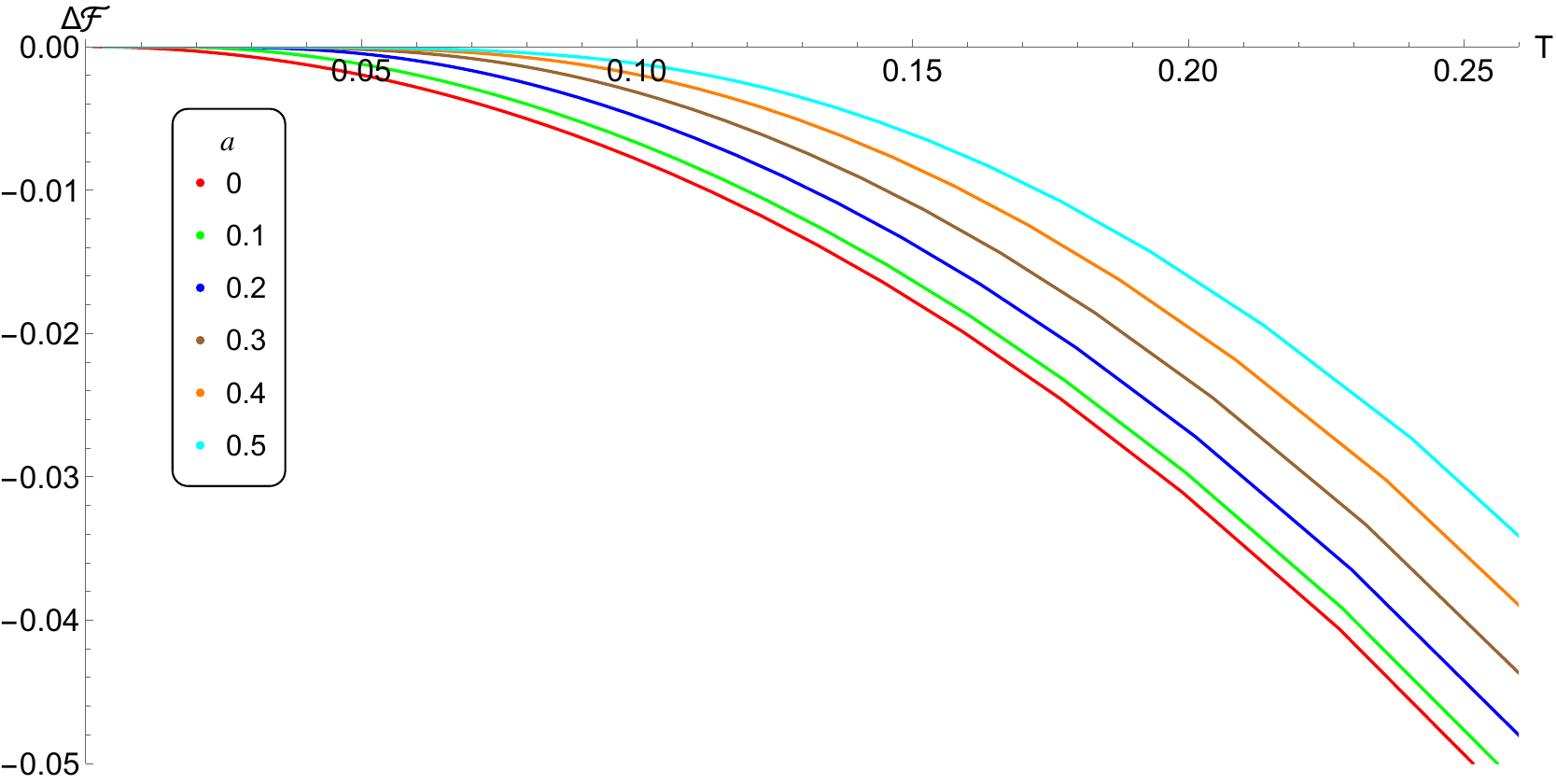}
\caption{\small Free energy difference $\Delta \mathcal{F}$ plotted against the black hole temperature $T$ for varying $a$ keeping $q=0$ fixed. }
\label{FvsTcase1a}
\end{minipage}
\hspace{0.4cm}
\begin{minipage}[b]{0.5\linewidth}
\centering
\includegraphics[width=2.8in,height=2.3in]{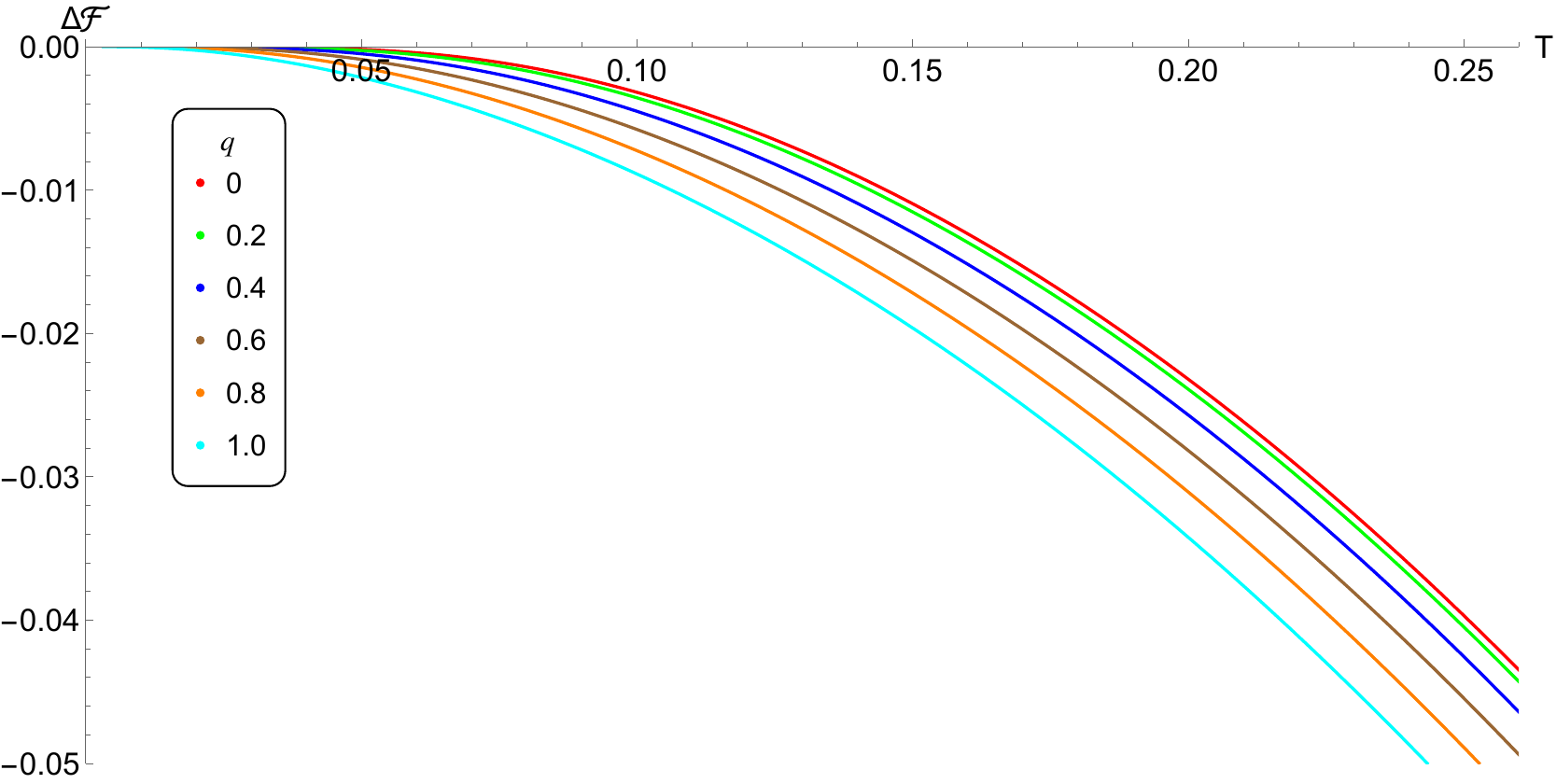}
\caption{\small Free energy difference $\Delta \mathcal{F}$ plotted against the black hole temperature $T$ for varying $q$ keeping $a=0.3$ and $\alpha=0.5$ fixed.}
\label{FvsTcase1q}
\end{minipage}
\end{figure}

\begin{figure}[h!]
\begin{minipage}[b]{0.5\linewidth}
\centering
\includegraphics[width=2.8in,height=2.3in]{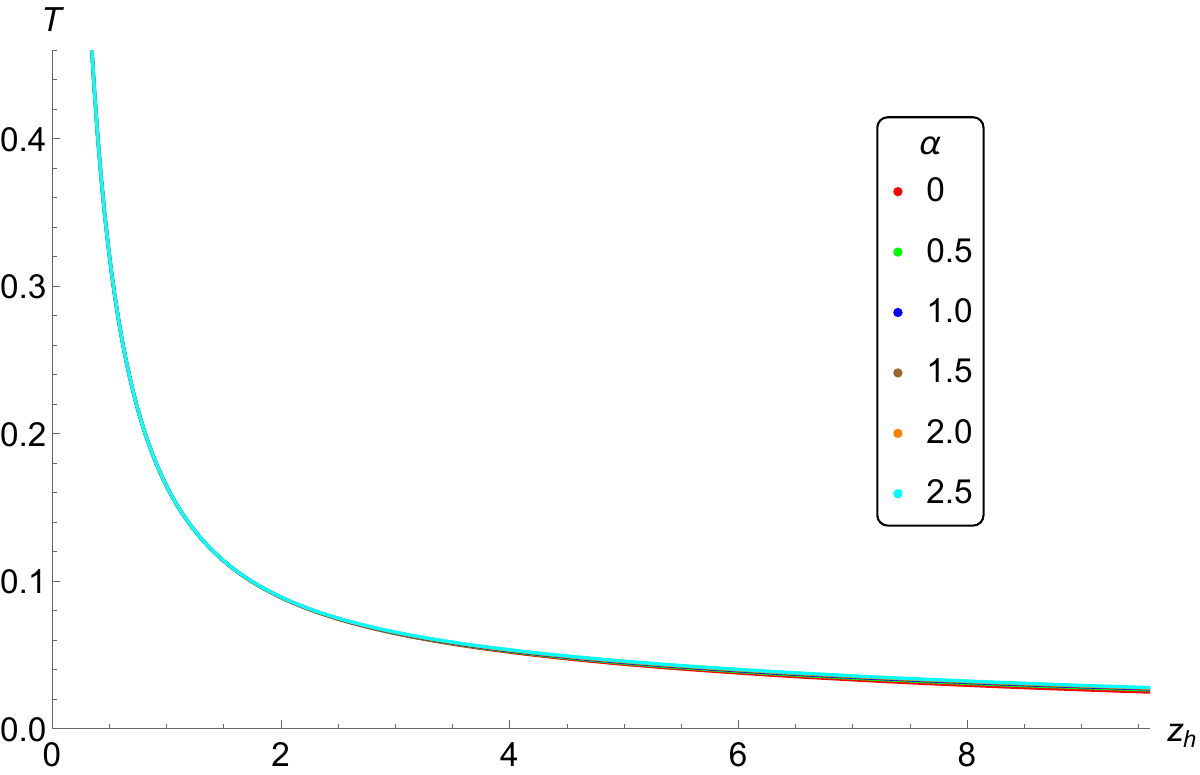}
\caption{ \small The nature of black hole temperature, $T$, plotted against the inverse horizon radius $z_h$ for varying $\alpha$ keeping $a=0.3$ and $q=0.2$ fixed.}
\label{tvszhcase1-alpha}
\end{minipage}
\hspace{0.4cm}
\begin{minipage}[b]{0.5\linewidth}
\centering
\includegraphics[width=2.8in,height=2.3in]{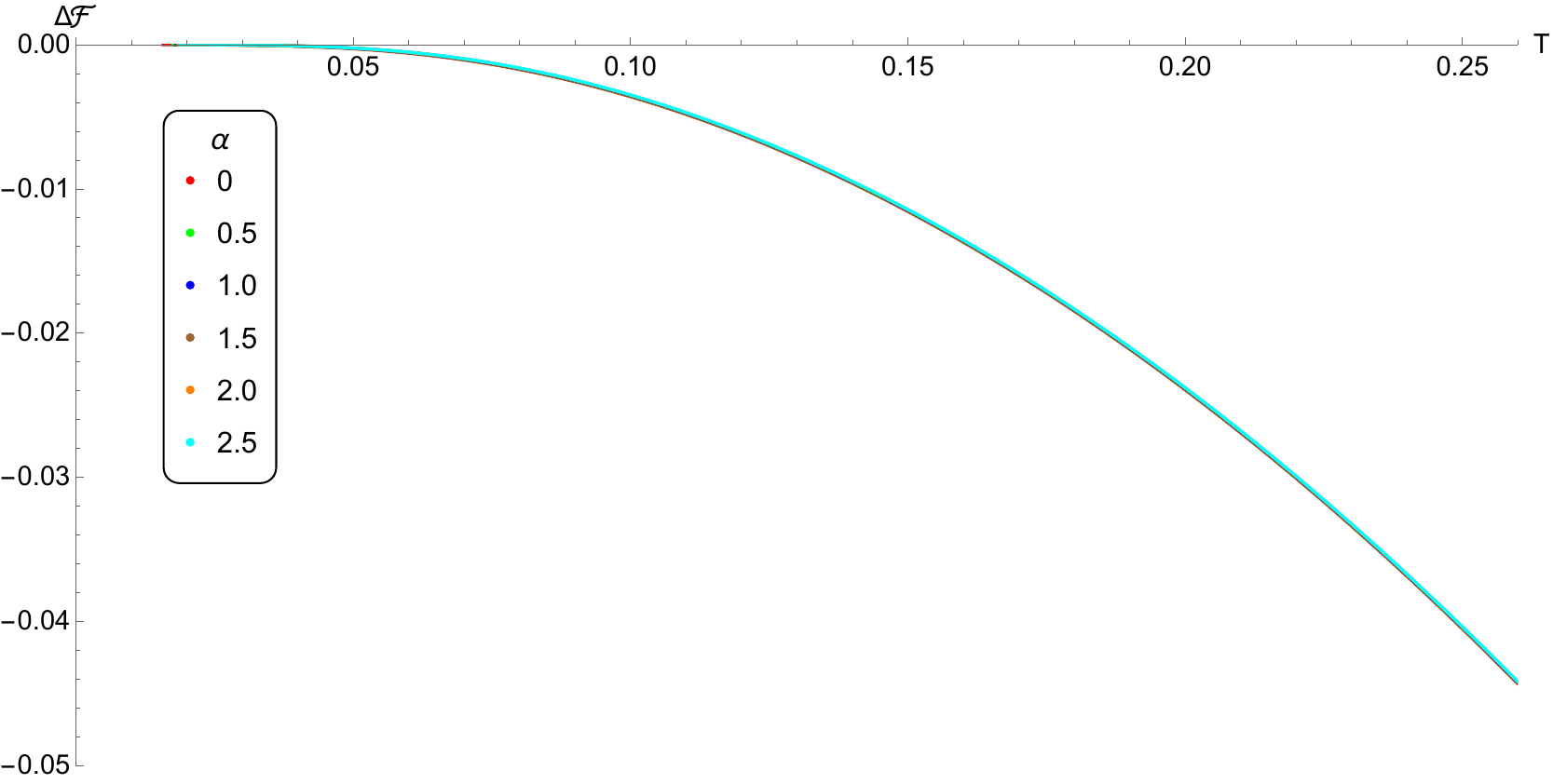}
\caption{\small Free energy difference $\Delta \mathcal{F}$ plotted against the black hole temperature $T_{HP}$ for varying $\alpha$ keeping $a=0.3$ and $q=0.2$ fixed.}
\label{FvsT_case-1_alpha}
\end{minipage}
\end{figure}

Examining their free energy patterns is necessary to delve deeper into the global thermodynamic stability of the aforementioned hairy black hole phases. In canonical ensemble, the Helmholtz free energy $\mathcal{F}$, in its differential form, is associated with the black hole entropy as follows:
\begin{eqnarray}
& & d\mathcal{F}=-S_{BH} \ d T\,.
\label{dFSdT}
\end{eqnarray}
Using the above equation, we can calculate the difference of the free energy between the black hole and thermal-AdS phases, the expression for which is given as follows\footnote{For the computation of free energy difference between black hole phases, the inverse horizon radius ($z_h$) for the thermal-AdS is considered to be at infinity.  }:
\begin{eqnarray}
& & \Delta\mathcal{F} =-\int S_{BH}d T = - \int^{z_h}_{z_\Lambda=\infty} S_{BH} \frac{d T}{d z_h} dz_h \,.
\label{Gibbsfreeenergy}
\end{eqnarray}
In Fig.~\ref{FvsTcase1a}, the Helmholtz free energy profile of the hairy black hole phase against its Hawking temperature is shown for various values of $a$. The colour scheme used here is the same as in Fig.~\ref{Tvszhcase1a}. Similarly, in Fig. \ref{FvsTcase1q} , we show plot the free energy against $T$ for different values $q$ (when $a=0.3$ and $\alpha=0.5$). 

Similar to the previous cases, in Fig.~\ref{tvszhcase1-alpha}, the temperature profile is shown for different values of Born-Infeld parameter $\alpha$ with $a=0.3$ and $q=0.2$ fixed. Here, as well, it is observed that only one thermodynamically stable phase is favourable at all temperatures for all values of $\alpha$ with no strong dependence of temperature on the strength of $\alpha$. In Fig.~\ref{FvsT_case-1_alpha}, the free energy is plotted for different values $\alpha$ (when $a=0.3$ and $q=0.2$). We find that the free energy does not change its sign for all finite values of $a$, $q$, and $\alpha$, confirming the existence of only a single stable black hole phase. The phase of a stable black hole reaches an extremal state at a certain horizon radius, denoted as $z_{h}^{ext}$. We can verify that the free energy of this stable phase is always lower than that of thermal-AdS. This observation is strikingly similar to the characteristics of a charged BTZ black hole. In the case of a non-hairy charged BTZ black hole (with Born-Infeld potential), the extremal horizon radius can be determined from Eq.~(\ref{tempcase1a0}), and it occurs when $z_{h}^{ext}=\frac{\sqrt{e^{4 \alpha }-1}}{\sqrt{\alpha } q}$. However, for a hairy black hole, the value of $z_{h}^{ext}$ escalates with an increase in $a$.

\section{\label{sec:f=1} Solution of the hairy black hole with form factor $A(z)=-a^2 z^2$ }

In this section, we investigate the geometry and thermodynamics of black hole solution with the coupling function and the form factor given by $f(z)$ = $e^{-A(z)} \sqrt{\alpha ^2 q^4 z^4+1}$ and $A(z)=-a^2 z^2$, respectively. Thus, from the Eq.~\eqref{phisol}, the scalar field solution is obtained as follows:
\begin{equation}
\phi(z)=a z \sqrt{2 a^2 z^2+3}+\frac{3 \left(\log (3)-2 \log \left(\sqrt{2 a^2 z^2+3}-\sqrt{2} a z\right)\right)}{2 \sqrt{2}}.
\label{phivalue2}
\end{equation}
Note that the scalar field vanishes as $a\to0$. For this particular coupling as well, it is observed that the scalar field maintains its regularity, finiteness, and stability in all regions beyond the horizon. Using Eq. \eqref{electricfield}, the expression for the gauge field can be found as,
\begin{equation}
    \begin{aligned}
   B_t(z)&=\frac{q}{2}  \left(e^{\frac{2 a^2}{\alpha  q^2}} \left(\text{Ei}\left(2 a^2 \left(-z^2-\frac{1}{q^2 \alpha }\right)\right)-\text{Ei}\left(2 a^2 \left(-{z_h}^2-\frac{1}{q^2 \alpha }\right)\right)\right)\right) \\
   &+\frac{q}{2} \left(\text{Ei}\left(-2 a^2 {z_h}^2\right)-\text{Ei}\left(-2 a^2 z^2\right)\right).
   \end{aligned}
\end{equation}
When the hairy parameter $a$ vanishes, the gauge field reduces to Born-Infeld potential,
\begin{equation}
   B_t(z)\big\rvert_{a \rightarrow0}= -\frac{q}{2} \log \left(1+\frac{z^2-{z_h}^2}{{z_h}^2 \left(\alpha  q^2 z^2+1\right)}\right) .
\end{equation}
Now, in a similar fashion, we can obtain the blackening function $g(z)$ using Eq. \eqref{g(z)eqn} as follows:
\begin{equation}
    \begin{aligned}
         g(z) &=1+ \frac{q^2\left(e^{a^2 z^2} \text{Ei}\left(-2 a^2 z^2\right)-\text{Ei}\left(-a^2 z^2\right)-\log(2)-e^{\frac{a^2}{\alpha q^2}}\text{Ei}\left(-\frac{a^2}{q^2 \alpha }\right)\right)}{4 a^2}\\
         & +\frac{q^2 e^{\frac{a^2}{\alpha  q^2}} \left(\text{Ei}\left(a^2 \left(-z^2-\frac{1}{q^2 \alpha }\right)\right)-e^{a^2 \left(\frac{1}{\alpha  q^2}+z^2\right)} \text{Ei}\left(2 a^2 \left(-z^2-\frac{1}{q^2 \alpha }\right)\right)\right)}{4 a^2}\\
         & -\frac{(-1 + e^{a^2 z^2})q^2 \left(e^{a^2 {z_h}^2} \text{Ei}\left(-2 a^2 {z_h}^2\right)-\text{Ei}\left(-a^2 {z_h}^2\right)-e^{\frac{a^2}{\alpha  q^2}}\text{Ei}\left(-\frac{a^2}{\alpha  q^2}\right)\right)}{4 a^2(e^{a^2 {z_h}^2}-1)}\\
         & - \frac{(-1 + e^{a^2 z^2}) q^2 e^{\frac{a^2}{\alpha  q^2}}\left(e^{\frac{a^2}{\alpha  q^2}} \text{Ei}\left(-\frac{2 a^2}{q^2 \alpha }\right)+\text{Ei}\left(a^2 \left(-{z_h}^2-\frac{1}{q^2 \alpha }\right)\right)\right)}{4 a^2(e^{a^2 {z_h}^2}-1)} \\
         & +\frac{(-1 + e^{a^2 z^2}) q^2 e^{\frac{a^2}{\alpha  q^2}}e^{a^2 \left(\frac{1}{\alpha  q^2}+{z_h}^2\right)} \text{Ei}\left(2 a^2 \left(-{z_h}^2-\frac{1}{q^2 \alpha }\right)\right)}{4 a^2(e^{a^2 {z_h}^2}-1)}\\
         & +\frac{q^2 e^{\frac{2a^2}{\alpha  q^2}}\text{Ei}\left(-\frac{2 a^2}{q^2 \alpha }\right)}{4 a^2}+\frac{(-1 + e^{a^2 z^2}) \left(q^2\log(2)-4 a^2\right)}{4 a^2(e^{a^2 {z_h}^2}-1)}\,.
        \label{g(z)2}
    \end{aligned}
\end{equation}
Here, $Ei$ denotes the exponential integral function.
Note that, as the hairy parameter and, subsequently, the scalar field is turned off, this expression reduces to the expression of the standard charged BTZ black hole with Born-Infeld potential as follows:
\begin{equation}
    \begin{aligned}
      g(z)\big\rvert_{a \rightarrow0}&=\frac{ \left(2 \alpha  q^2 z^2 \log \left(\frac{z}{{z_h}}\right)-\left(\alpha  q^2 z^2+1\right) \log \left(\alpha  q^2 z^2+1\right)\right)}{4 \alpha}\\
      & +\frac{\left({z_h}^2-z^2\right)}{{z_h}^2}+\frac{z^2 \left(\alpha  q^2 {z_h}^2+1\right) \log \left(\alpha  q^2 {z_h}^2+1\right)}{4 \alpha {z_h}^2} .
    \end{aligned}
\end{equation}
We can similarly calculate the expression for $V(z)$ using Eq.~(\ref{Vsol}).

%%%%%%%%%%%%%%%%%%%%%%%%%%%%%%
\begin{figure}[h!]
\begin{minipage}[b]{0.5\linewidth}
\centering
\includegraphics[width=2.8in,height=2.3in]{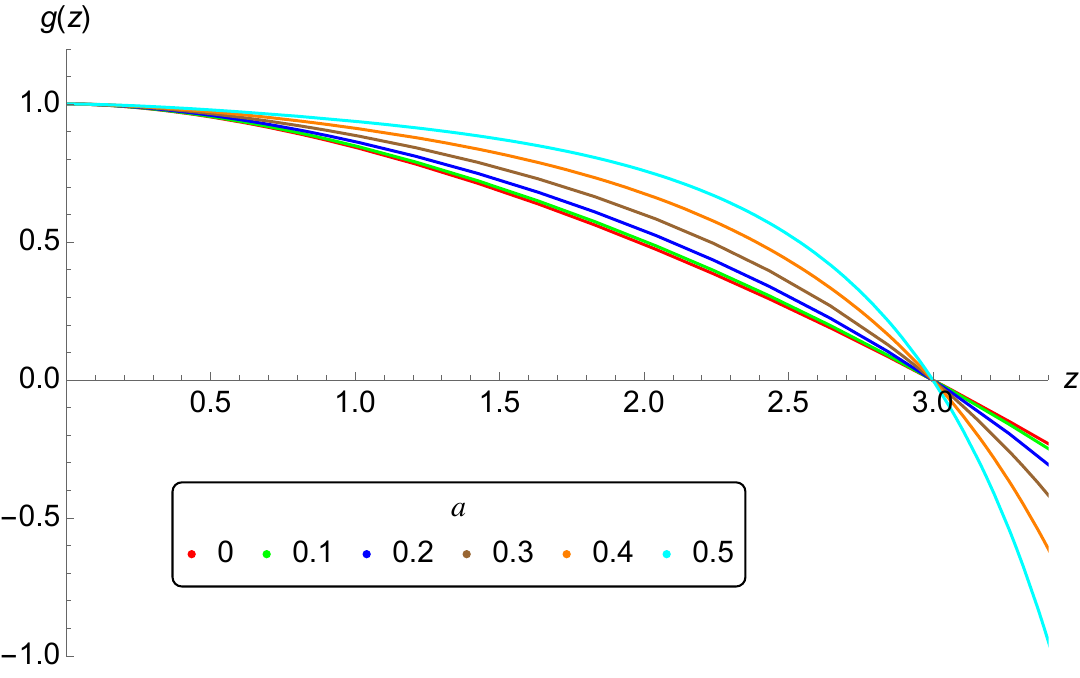}
\caption{ \small The radial profile of the blackening function for varying $a$ keeping $z_{h}=3.0$, $\alpha=0.5$, and $q=0.3$ fixed. }
\label{GvsZaF2qPt3alphaPt5}
\end{minipage}
\hspace{0.4cm}
\begin{minipage}[b]{0.5\linewidth}
\centering
\includegraphics[width=2.8in,height=2.3in]{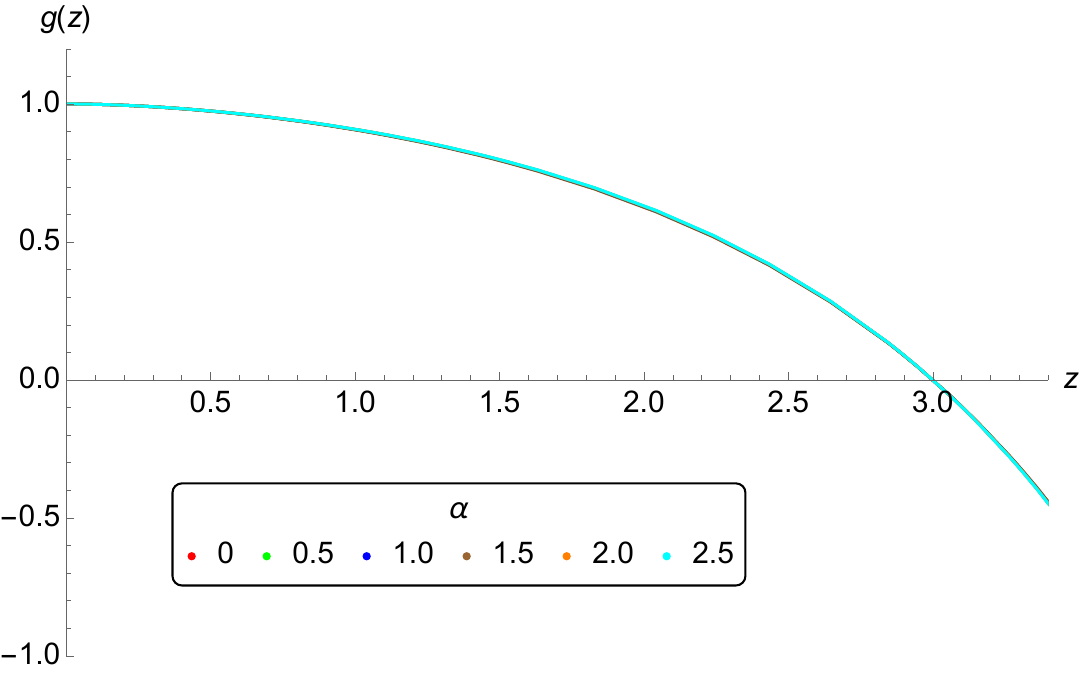}
\caption{\small The radial profile of the blackening function for varying $\alpha$ keeping $z_{h}=3.0$, $a=0.3$, and $q=0.2$ fixed. }
\label{GvsZalphaF2aPt3qPt2}
\end{minipage}
\end{figure}

\begin{figure}[h!]
\begin{minipage}[b]{0.5\linewidth}
\centering
\includegraphics[width=2.8in,height=2.3in]{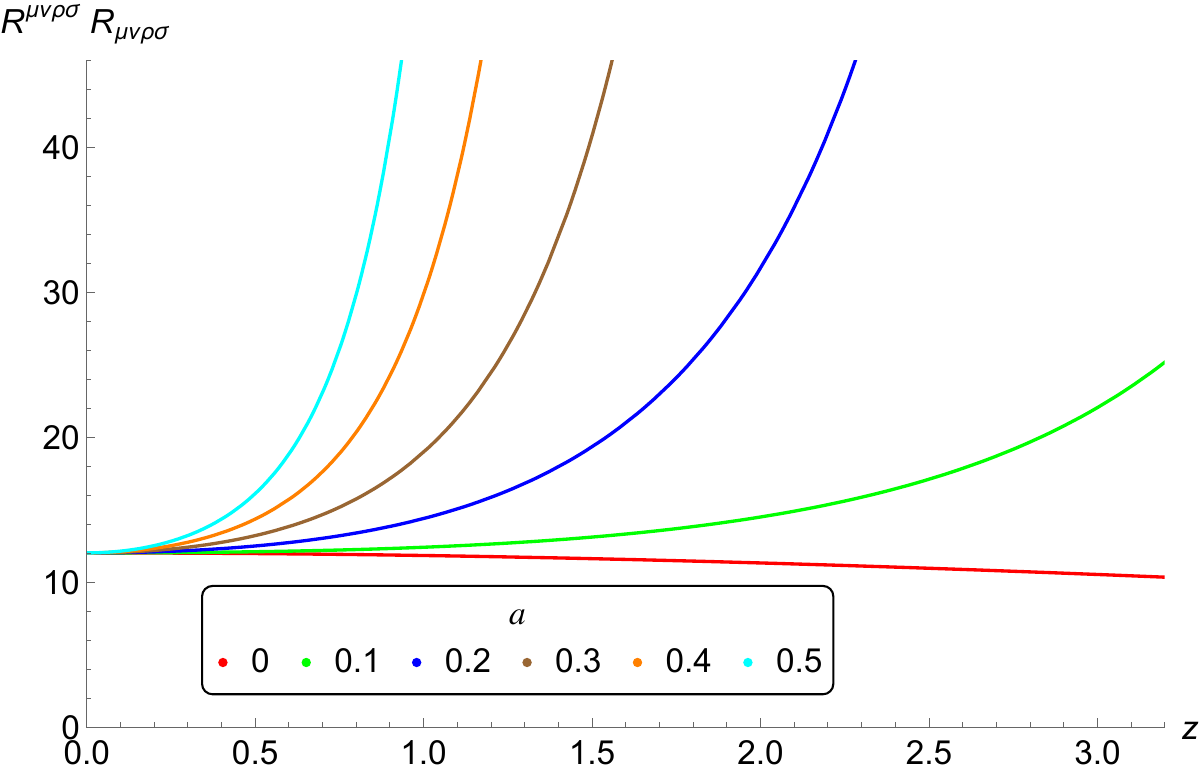}
\caption{ \small The nature of $R_{\mu\nu\rho\sigma}R^{\mu\nu\rho\sigma}$ for varying $a$ keeping $z_{h}=3.0$, $\alpha=0.5$, and $q=0.3$ fixed.}
\label{KSvsZaF2qPt3alphaPt5}
\end{minipage}
\hspace{0.4cm}
\begin{minipage}[b]{0.5\linewidth}
\centering
\includegraphics[width=2.8in,height=2.3in]{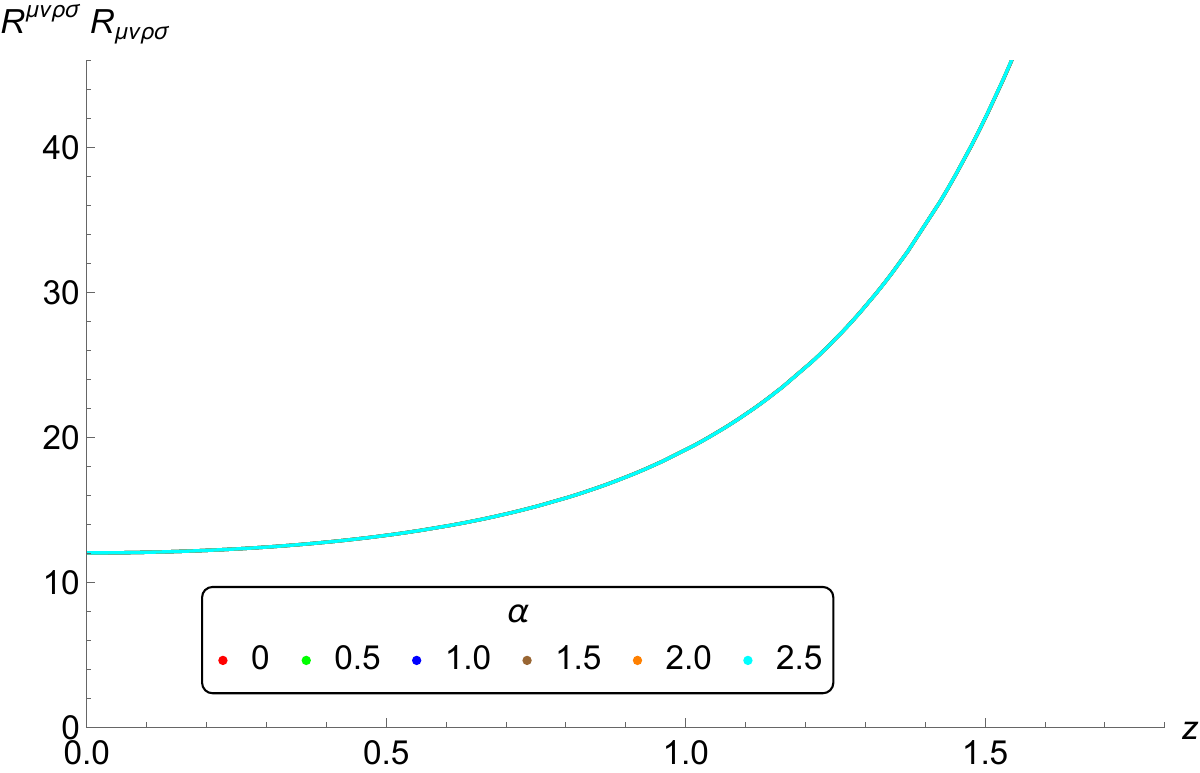}
\caption{\small The nature of $R_{\mu\nu\rho\sigma}R^{\mu\nu\rho\sigma}$ for varying $\alpha$ keeping $z_{h}=3.0$, $a=0.3$, and $q=0.2$ fixed. }
\label{KSvsZalphaF2aPt3qPt2}
\end{minipage}
\end{figure}
\begin{figure}[h!]

    \subfigure[]{\includegraphics[width=0.5\textwidth]{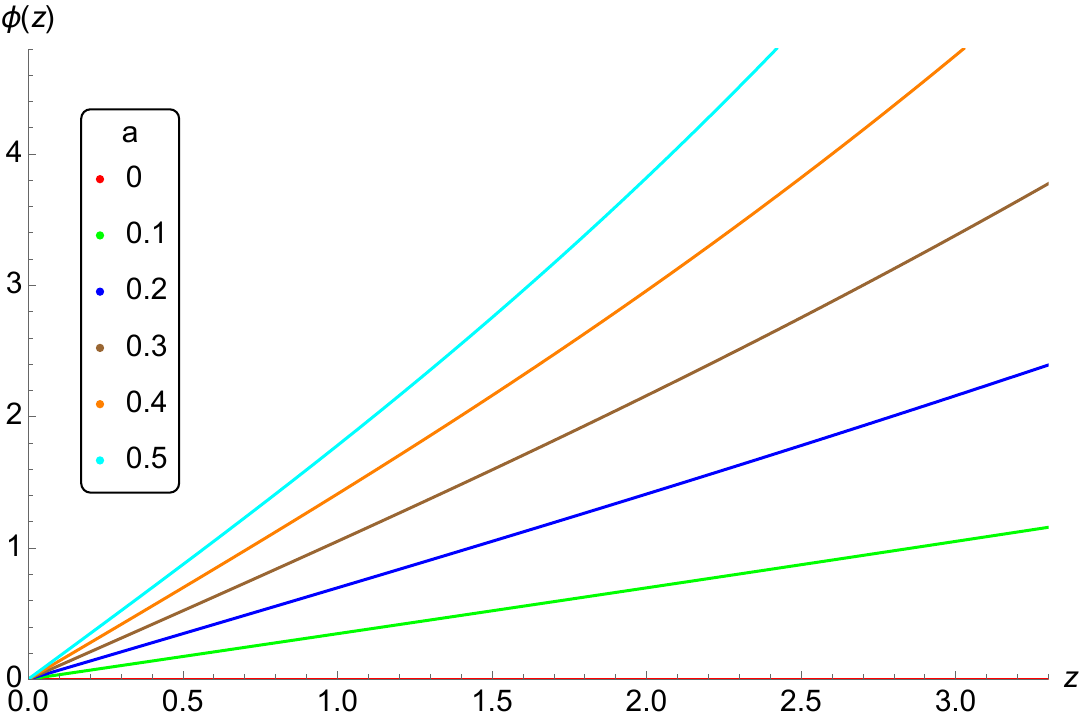}}
    \subfigure[]{\includegraphics[width=0.5\textwidth]{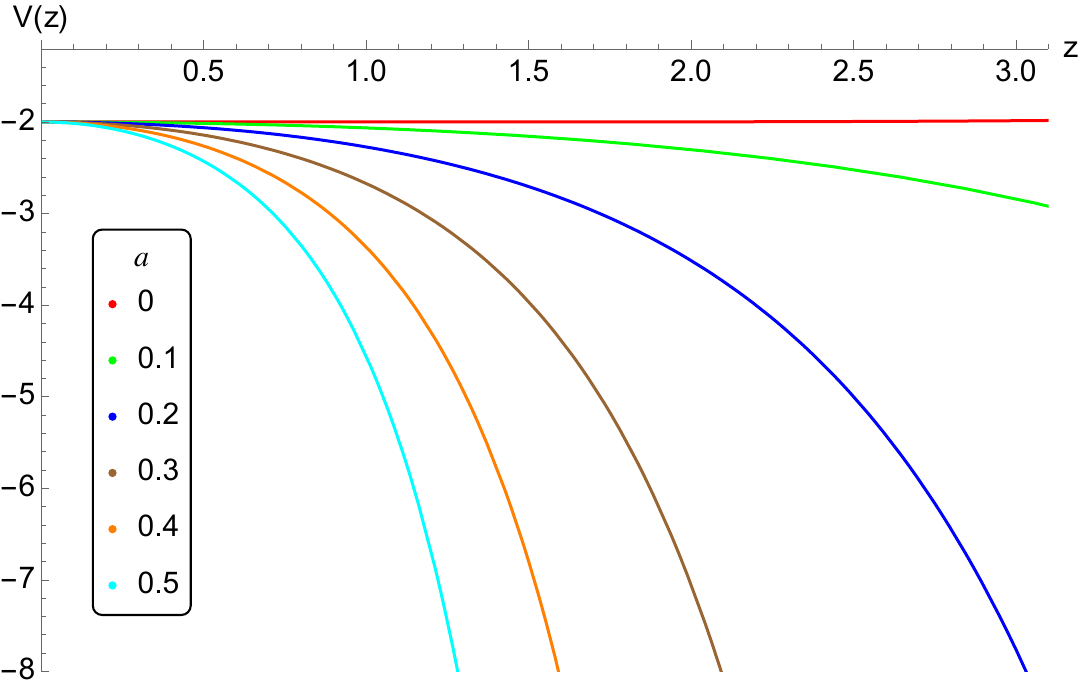}}
    \caption{The radial nature of the scalar field and its potential for varying $a$ keeping $z_{h}=3.0$, $\alpha=0.5$, and $q=0.3$ fixed.  }
    \label{phi_V_F2}
\end{figure}
%%%%%%%%%%%%%%%%%%%%%%%%%%%%%%
In Figs.~\ref{GvsZaF2qPt3alphaPt5} and \ref{GvsZalphaF2aPt3qPt2}, the radial profile of $g(z)$ is plotted against the inverse horizon radius $z_h$ for the different values of the hairy parameter $a$ and the BI non-linearity parameter $\alpha$, respectively. We have taken $z_h=3$ and $q=0.3$. At $z=z_h$, a change in the sign of $g(z)$ is observed, confirming the presence of the horizon for all values of $a$ and $\alpha$. Figs.~\ref{KSvsZaF2qPt3alphaPt5} and \ref{KSvsZalphaF2aPt3qPt2} show the radial profiles of the Kretschmann scalar $R_{\mu\nu\rho\sigma}R^{\mu\nu\rho\sigma}$ for different values of $a$ and $\alpha$, respectively. We can observe from the profile of the Kretschmann scalar that the geometry remains regular everywhere outside of the horizon except at $z=\infty$ or $r=0$. The Ricci scalar is also found to be regular and finite throughout the spacetime outside of the horizon. Also, note that no extra curvature singularities arise due to the introduction of the hairiness to the black hole compared to those present in the case of a non-hairy charged BTZ black hole. But unlike the non-hairy uncharged BTZ case, the uncharged hairy black hole in our model possesses a curvature singularity. Also, while the intensity of the singularity is directly proportional to the strength of the scalar field parameter $a$, no such strong dependence is observed on the value of $\alpha$.

From Fig.~\ref{phi_V_F2}, we can observe that the scalar field is a function of $z$ and $a$. It is finite and real everywhere except at the centre of the black hole and vanishes only at the asymptotic boundary. This suggests the presence of a well-behaved hairy black hole solution in three dimensions. Moreover, the potential approaches a constant upper limit, $V(z)|_{z\rightarrow 0} = 2\Lambda$, at the AdS boundary. The radial profile of the potential for different values of hair parameter $a$ is shown in Fig.~\ref{phi_V_F2}.

Now, we can move on to discuss the thermodynamics of this black hole solution. Using Eq.~\eqref{tempandentropy}, the temperature of the black hole is found to be:
\begin{equation}
    \begin{aligned}
        T & = \frac{z_h q^2 e^{a^2 z_h^2 } e^{\frac{a^2}{\alpha  q^2}} \left(e^{\frac{a^2}{\alpha  q^2}} \left(\text{Ei}\left(-\frac{2 a^2}{q^2 \alpha }\right)-\text{Ei}\left(2 a^2 \left(-z_{h}^2-\frac{1}{q^2 \alpha }\right)\right)\right)+\text{Ei}\left(a^2 \left(-z_{h}^2-\frac{1}{q^2 \alpha }\right)\right)\right)}{8 \pi  \left(e^{a^2 z_{h}^2}-1\right)}\\
        &+ \frac{z_h e^{a^2 z_h^2}\left(4 a^2+q^2\left(\text{Ei}\left(-2 a^2 z_{h}^2\right)-\text{Ei}\left(-a^2 z_{h}^2 \right)-\log(2)-e^{\frac{a^2}{\alpha  q^2}}\text{Ei}\left(-\frac{a^2}{q^2 \alpha }\right)\right)\right)}{8 \pi  \left(e^{a^2 z_{h}^2}-1\right)} .
    \end{aligned}
\end{equation}
When the hairy parameter vanishes, the above expression for temperature simplifies to the typical BTZ-like temperature expression with Born-Infeld potential, i.e.,
\begin{equation}
  T\big\rvert_{a \rightarrow0} = \frac{1}{2 \pi z_h}-\frac{\log \left(1+\alpha  q^2 z_{h}^2\right)}{8 \pi  \alpha z_h}\,,
\label{temp2a0}  
\end{equation}
and in the limit ($a\to0$, $q \to 0$), it smoothly reduces to the temperature expression for the standard BTZ solution,
\begin{equation}
  T\big\rvert_{a \rightarrow0, q\rightarrow0 } = \frac{1}{2 \pi z_h}\,.
\label{temp2a0q0}  
\end{equation}
In this case, as well, it is interesting to observe that for particular values of the charge parameter $q$, the temperature of the black hole vanishes as opposed to the case of uncharged black holes. Such black holes with vanishing temperatures are known as `extremal' black holes.

\begin{figure}[h!]
\begin{minipage}[b]{0.5\linewidth}
\centering
\includegraphics[width=2.8in,height=2.3in]{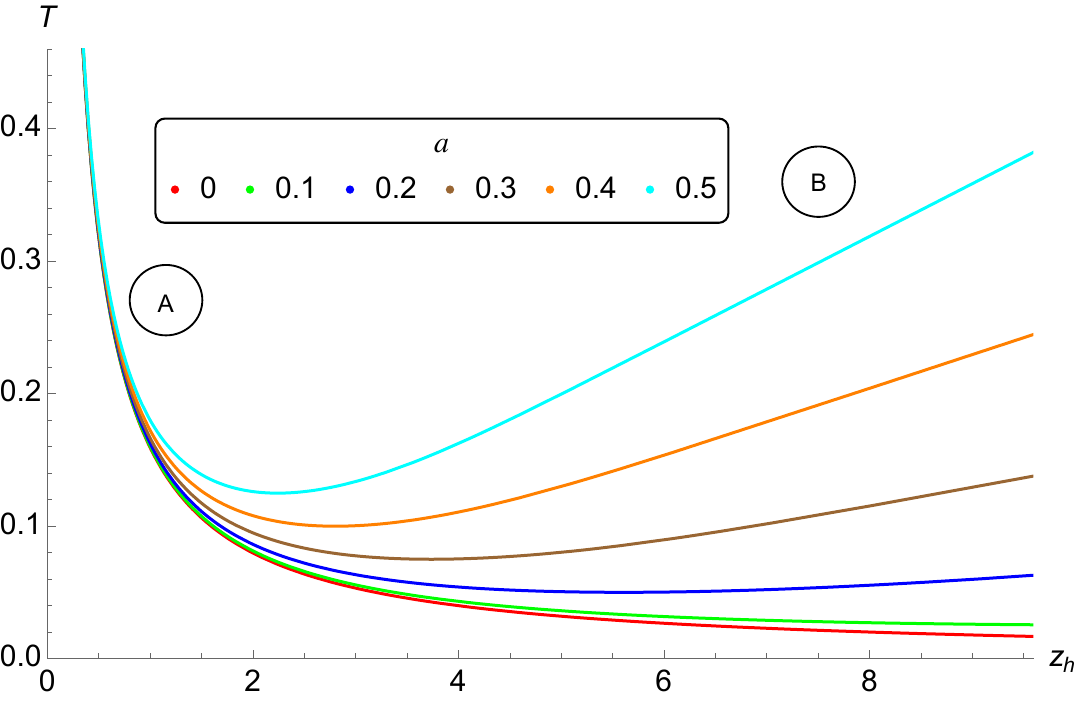}
\caption{ \small The nature of black hole temperature $T$ plotted against the inverse horizon radius $z_h$ for varying $a$ keeping $q=0$ and $\alpha=0.5$ fixed.}
\label{tvszhcase2-a}
\end{minipage}
\hspace{0.4cm}
\begin{minipage}[b]{0.5\linewidth}
\centering
\includegraphics[width=2.8in,height=2.3in]{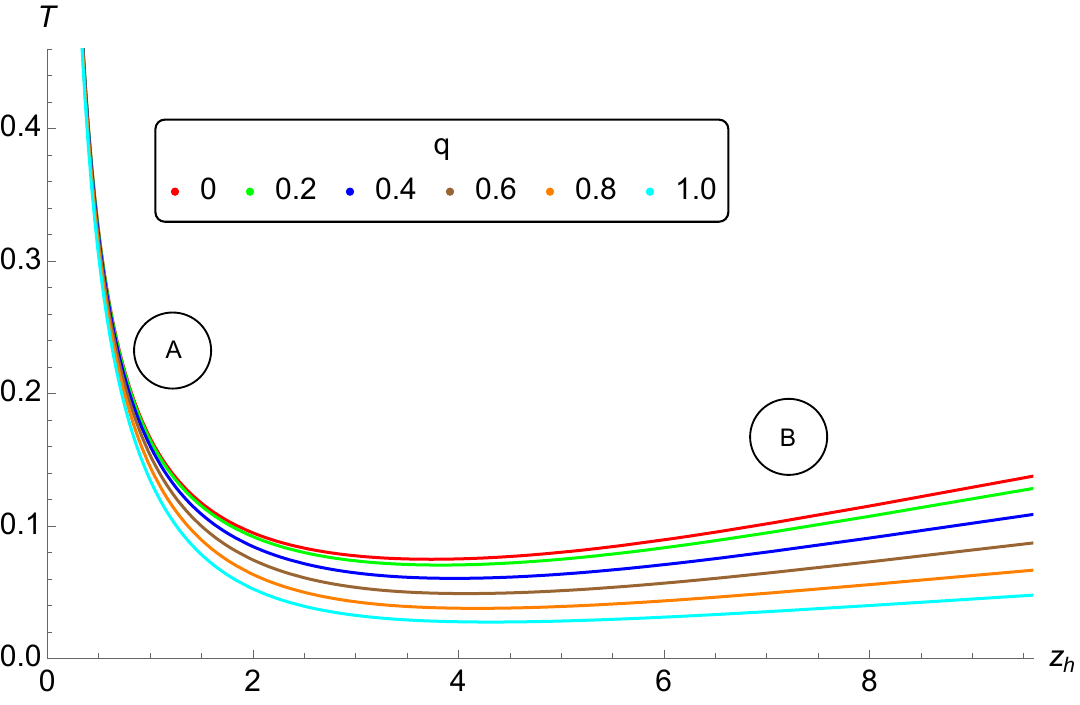}
\caption{\small The nature of black hole temperature $T$ plotted against the inverse horizon radius $z_h$ for varying $q$ keeping $a=0.3$ and $\alpha=0.5$ fixed.}
\label{tvszhcase2-q}
\end{minipage}
\end{figure}

In Fig.~\ref{tvszhcase2-a}, the Hawking temperature of the black hole against the inverse horizon radius $z_h$ is plotted for different values of the hairy parameter $a$. Here, we have fixed $q=0$. When $a=0$, the black hole has only one stable branch corresponding to the non-hairy charged BTZ black hole. It is interesting to note that for all non-vanishing values of $a$, above some critical temperature, there exist two black hole phases: a `small black hole phase' (unstable) with negative specific heat and a `large black hole phase' (stable) with positive specific heat. The stable and unstable phases in all figures are denoted by $\textcircled{A}$ and $\textcircled{B}$, respectively. With the increase in inverse horizon radius $z_h$, the temperature for the large (stable) black hole phase decreases while it increases for the small (unstable) black hole phase. It is apparent that both phases of the black hole can exist only above a certain temperature for all non-zero $a$. Thus, below this minimum temperature ($T_{min}$), no large or small black hole phases occur, and the thermal-AdS remains the only viable phase. This evinces the possibility of phase transition between the small/large phases of the black hole to the thermal-AdS. As we will see shortly, there is indeed a Hawking/Page phase transition between the large black hole and thermal-AdS phases.

%%%%%%%%%%%%%%%%%%%%%%%%%%%%%%

\begin{figure}[h!]
\begin{minipage}[b]{0.5\linewidth}
\centering
\includegraphics[width=2.8in,height=2.3in]{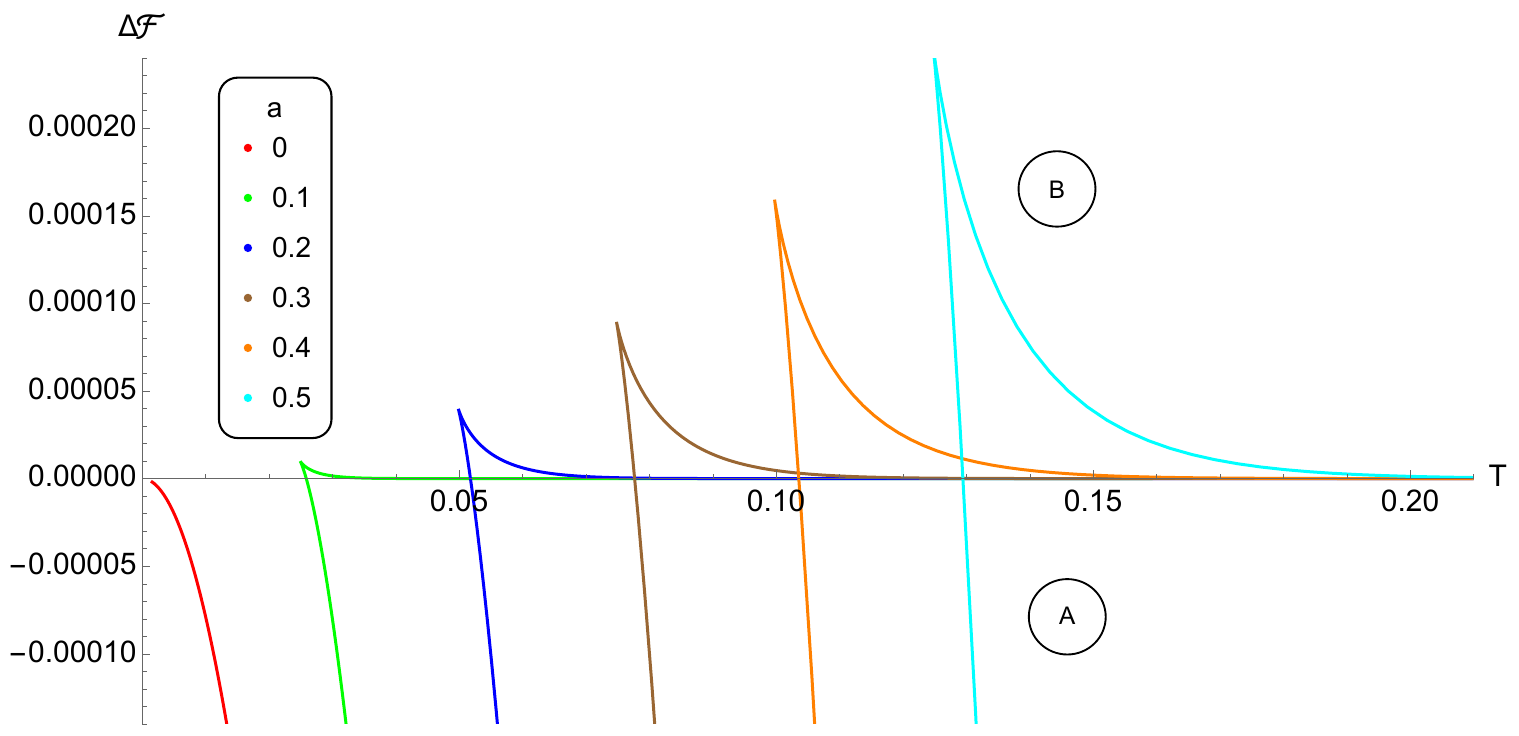}
\caption{ \small Free energy difference $\Delta \mathcal{F}$ plotted against the black hole temperature $T$ for varying $a$ keeping $q=0$ and $\alpha=0.5$ fixed. }
\label{FvsT_case-2_a}
\end{minipage}
\hspace{0.4cm}
\begin{minipage}[b]{0.5\linewidth}
\centering
\includegraphics[width=2.8in,height=2.3in]{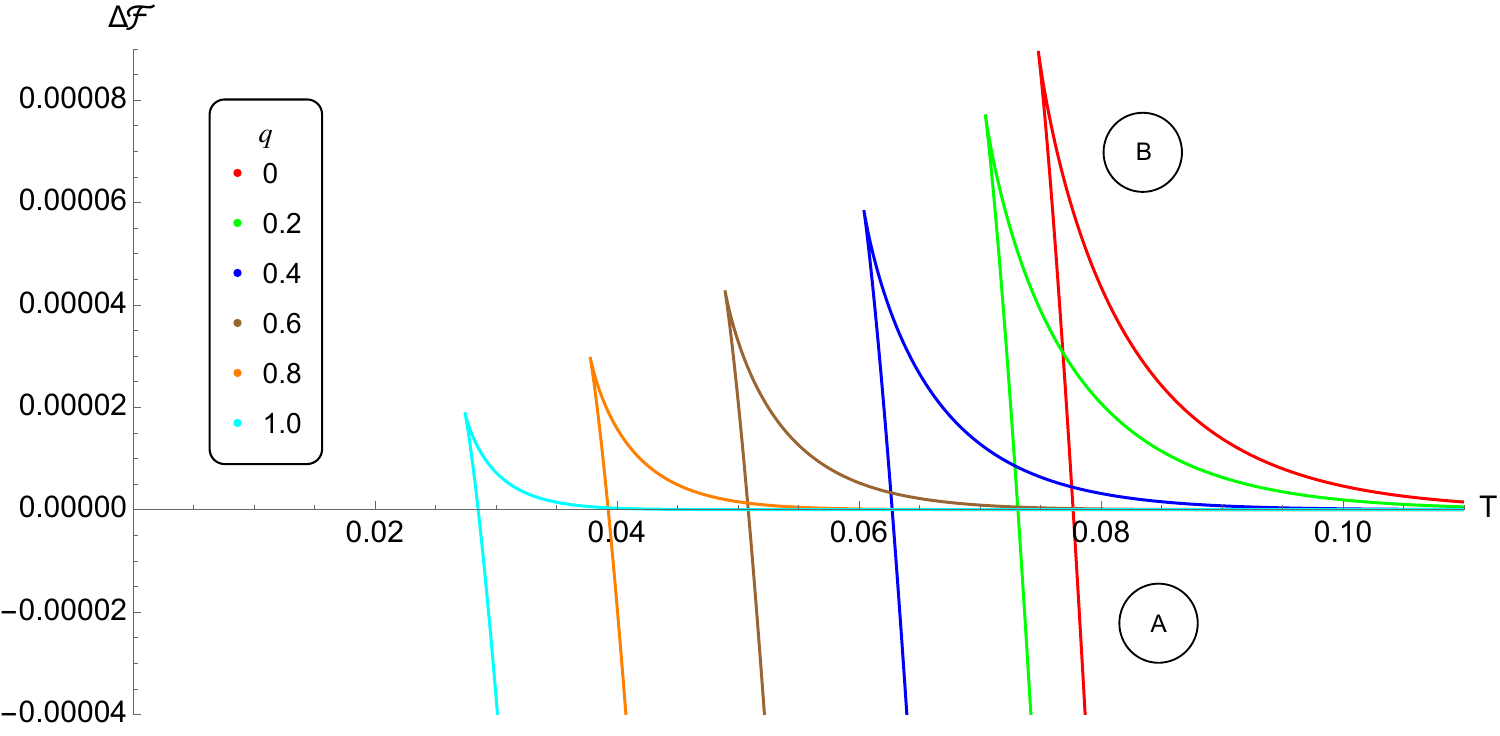}
\caption{\small Free energy difference $\Delta \mathcal{F}$ plotted against the black hole temperature $T$ for varying $q$ keeping $a=0.3$ and $\alpha=0.5$ fixed.}
\label{FvsT_case-2_q}
\end{minipage}
\end{figure}

\begin{figure}[h!]
\begin{minipage}[b]{0.5\linewidth}
\centering
\includegraphics[width=2.8in,height=2.3in]{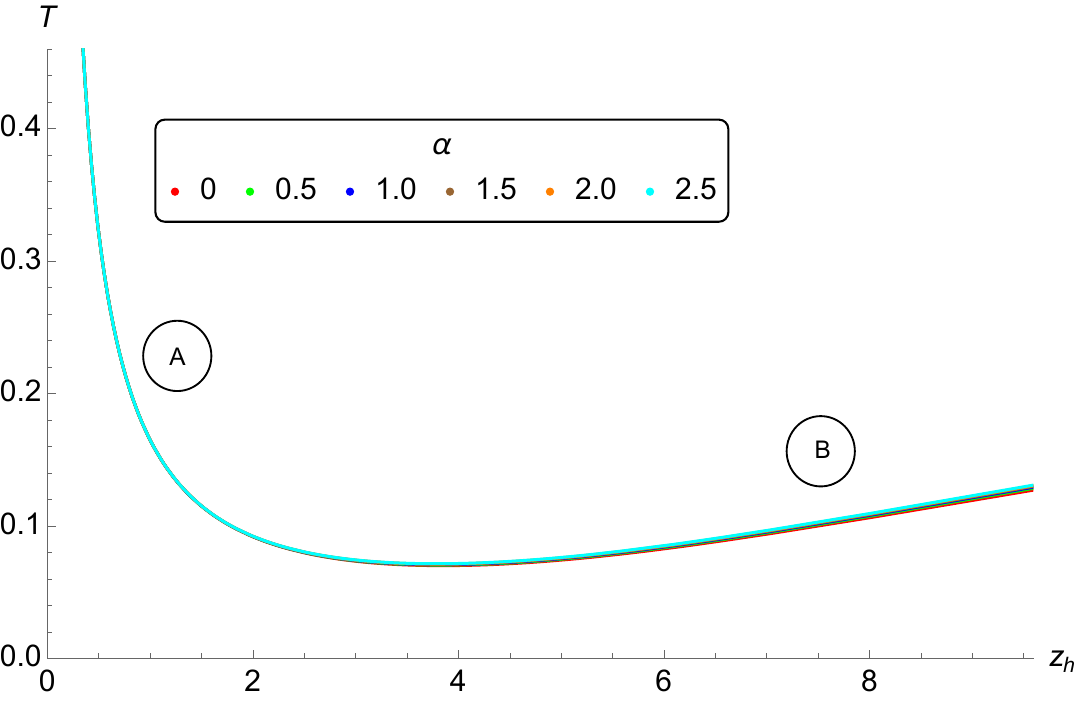}
\caption{ \small The nature of black hole temperature, $T$, plotted against the inverse horizon radius $z_h$ for varying $\alpha$ keeping $a=0.3$ and $q=0.2$ fixed.}
\label{tvszhcase2-alpha}
\end{minipage}
\hspace{0.4cm}
\begin{minipage}[b]{0.5\linewidth}
\centering
\includegraphics[width=2.8in,height=2.3in]{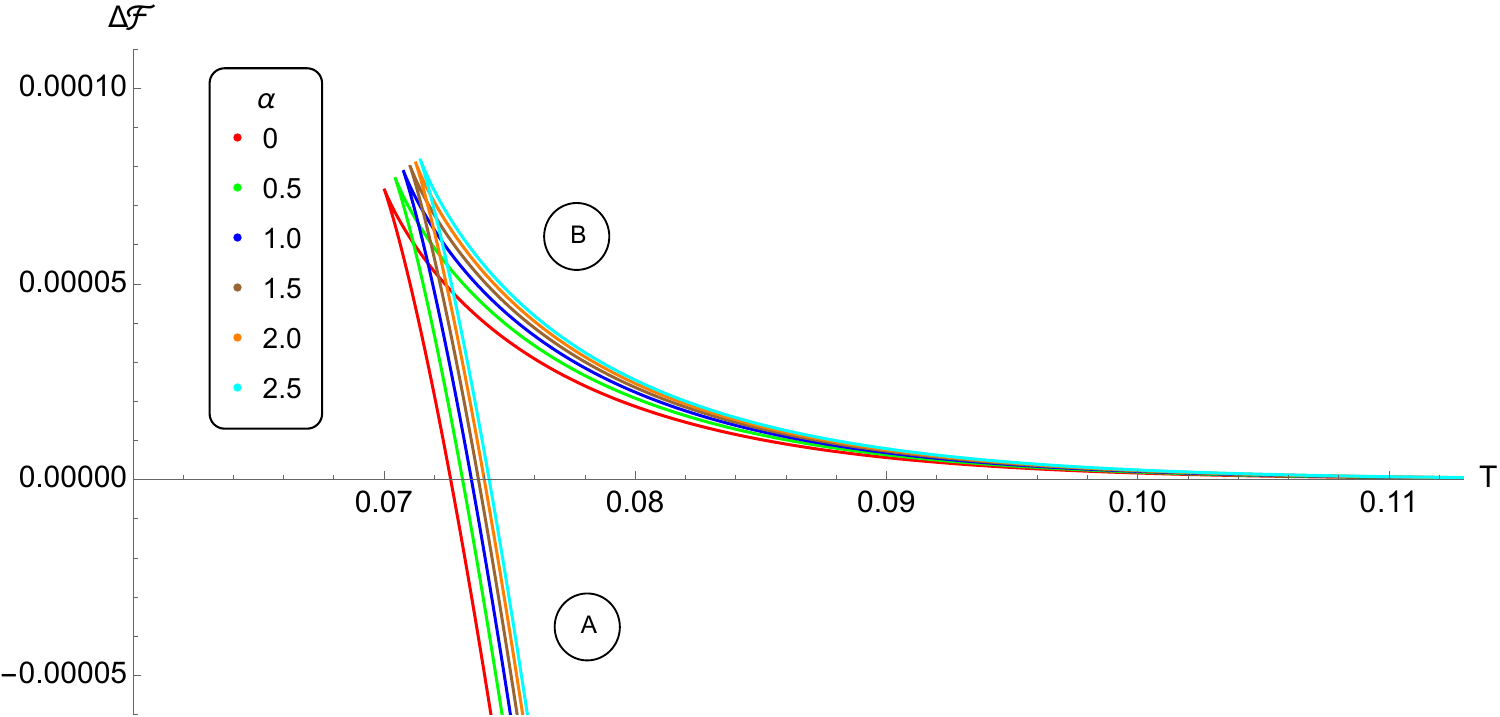}
\caption{\small Free energy difference $\Delta \mathcal{F}$ plotted against the black hole temperature $T_{HP}$ for varying $\alpha$ keeping $a=0.3$ and $q=0.2$ fixed.}
\label{FvsT_case-2_alpha}
\end{minipage}
\end{figure}
%%%%%%%%%%%%%%%%%%%%%%%%%%%%%%

Let us now look into the thermodynamics of the charged hairy black hole by switching on the parameter $q$. In Fig.~\ref{tvszhcase2-q}, temperature is plotted against the $z_h$ for different values of $q$ keeping $a=0.3$ and $\alpha=0.5$ fixed. In this scenario too, there exist two black hole phases (small and large) above some critical temperature, and as discussed in the previous case, the black hole can again undergo the Hawking/Page phase transition between a stable large black hole phase and the thermal-AdS phase. It is important to note that, for some value $q$, the black hole becomes extremal at a certain horizon radius, denoted as $z_{h}^{ext}$ with only one stable phase and no possibility of transition. In this case also, for a non-hairy charged BTZ black hole with Born-Infeld potential, the extremal horizon radius can be determined from Eq.~(\ref{temp2a0}), which is found to be equal to $z_{h}^{ext}=\frac{\sqrt{e^{4 \alpha }-1}}{\sqrt{\alpha } q}$. However, for a hairy black hole, the value of $z_{h}^{ext}$ escalates with an increase in $a$.

The above mentioned possibility of phase transition can be corroborated by studying the free energy as a function of Hawking temperature $T$. Fig.~\ref{FvsT_case-2_a} illustrates the nature of free energy difference with respect to the temperature of a black hole for different values of $a$, with $q=0$ fixed, following equivalent colour scheme as in Fig.~\ref{tvszhcase2-a}. The presence of two black hole phases and the Hawking/Page type of phase transition, for finite $a$, can be verified by looking at the multivaluedness and the change of sign of $\Delta \mathcal{F}$ plotted against $T$. For all non-vanishing values of $a$, there exists a minimum temperature $T_{min}$ below which no black hole phase exists. For temperature greater than $T_{min}$, two black hole phases are observed: a small black hole phase and a large black hole phase. Since the free energy of the large black hole phase is always smaller than that of the small black hole phase, it is more stable and, hence, thermodynamically favoured. Similarly, the free energy of the large black hole phase can be larger or smaller than the thermal-AdS phase, suggesting a phase transition between them. We define the temperature of Hawking/Page phase transition $T_{HP}$ to be the temperature at which the free energies of the large black hole phase and the thermal-AdS become equal, i.e., $\Delta \mathcal{F}=0$. This $T_{HP}$ is slightly above the $T_{min}$. It should be noted that the phase transition, in this case, is possible owing to the hairiness of our black hole solution, which can be contrasted with the lack of any phase transitions in the case of a non-hairy charged BTZ black hole. 

%%%%%%%%%%%%%%%%%%%%%%%%%%%%%%
\begin{figure}[h!]
\begin{minipage}[b]{0.5\linewidth}
\centering
\includegraphics[width=2.8in,height=2.3in]{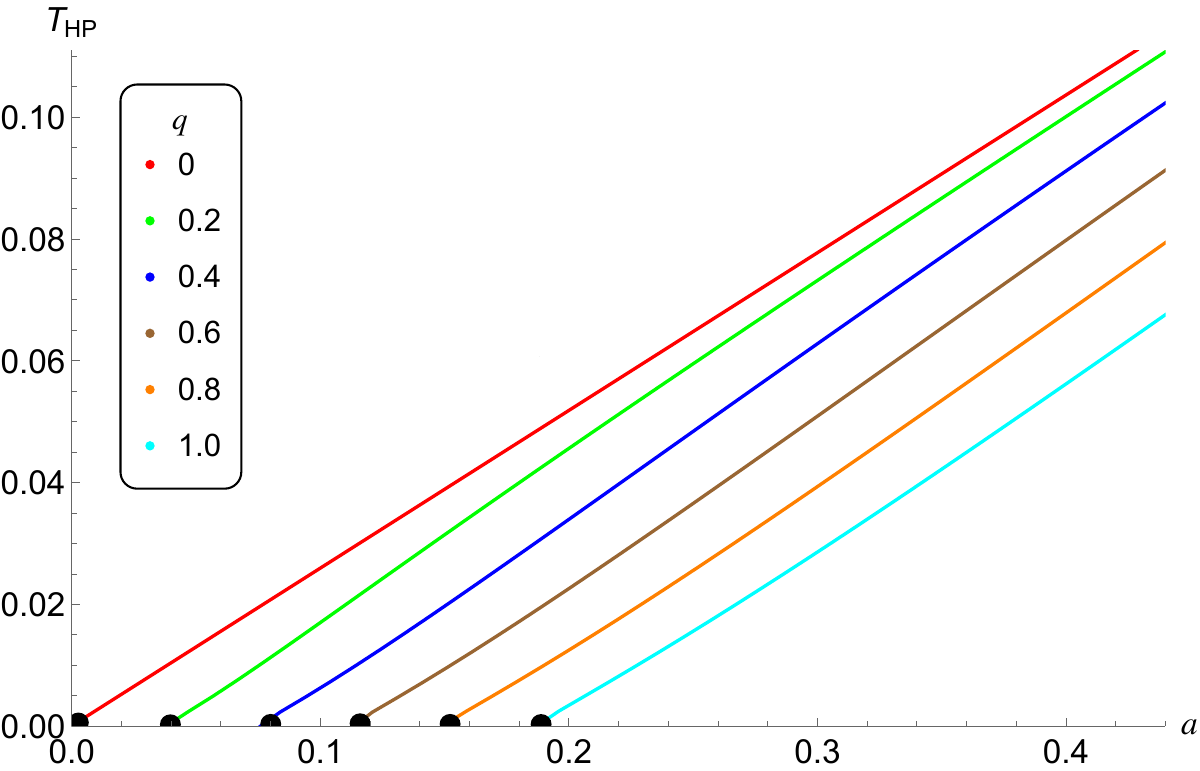}
\caption{ \small The Hawking/Page phase transition temperature $T_{HP}$ is plotted against $a$ for different values of $q$ when $\alpha$=0.5. }
\label{Crit T vs a}
\end{minipage}
\hspace{0.4cm}
\begin{minipage}[b]{0.5\linewidth}
\centering
\includegraphics[width=2.8in,height=2.3in]{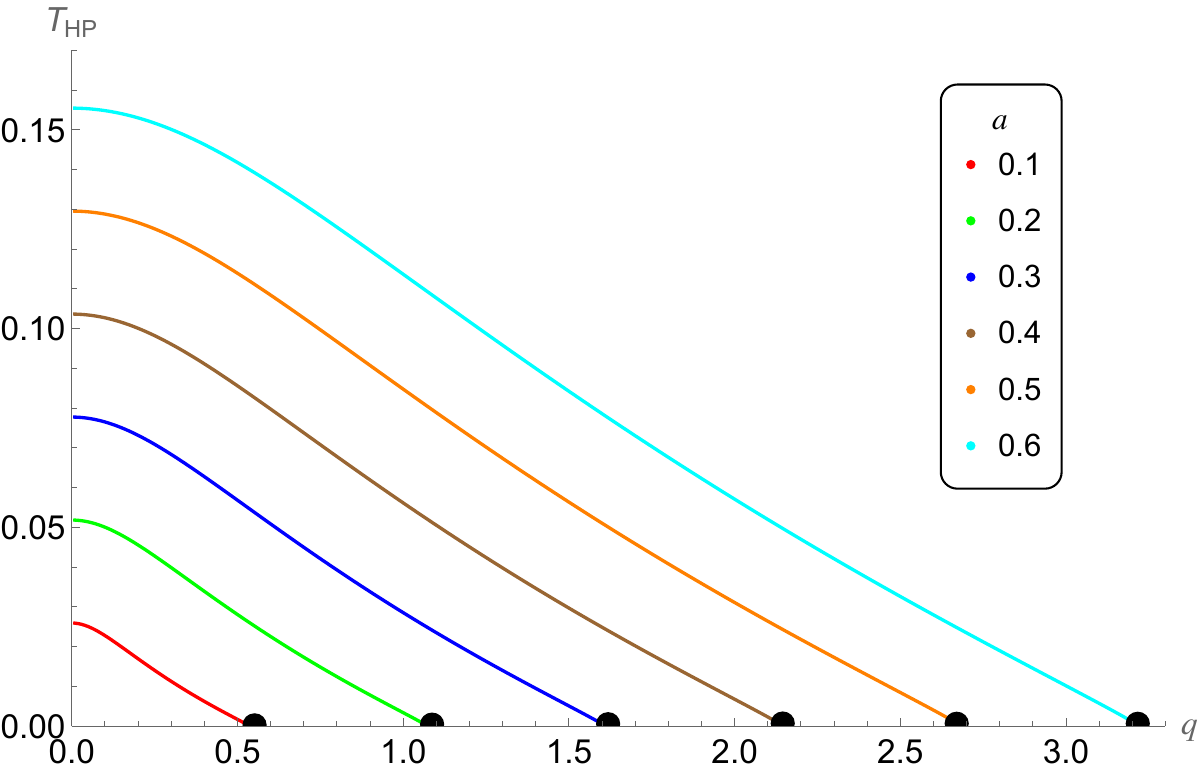}
\caption{\small The Hawking/Page phase transition temperature $T_{HP}$ is plotted against $q$ for various values of $a$ with $\alpha$=0.5.}
\label{Crit Tvsq}
\end{minipage}
\end{figure}
%%%%%%%%%%%%%%%%%%%%%%%%%%%%%%

Fig.~\ref{FvsT_case-2_q} illustrates the profile of free energy difference $\Delta \mathcal{F}$ against the Hawking temperature $T$ for various values of $q$, with $a=0.3$ and $\alpha=0.5$, following equivalent colour scheme as in Fig.~\ref{tvszhcase2-q}. For all considered values of $q$, again, Hawking/Page phase transition is observed above a certain critical temperature $T_{HP}$.

We can similarly analyze the dependence of temperature and free energy of the black hole on the Born-Infeld non-linearity parameter $\alpha$. In Figs.~\ref{tvszhcase2-alpha} and \ref{FvsT_case-2_alpha}, Hawking temperature and the free energy are plotted for different values of $\alpha$ keeping $a=0.3$ and $q=0.2$ fixed, respectively. Similar to previous cases, here also for all considered values of $\alpha$, two black hole phases are observed above $T_{min}$, with a possibility of phase transition between the large hairy black hole and thermal-AdS. Interestingly, we find that for a fixed $a$ and $q$, the Hawking/Page phase transition temperature increases as $\alpha$ increases, suggesting for higher values of $\alpha$, the black hole will exist at relatively higher temperatures.

Let us now study the dependence of the transition temperature $T_{HP}$ on $a$ and $q$. Fig.~\ref{Crit T vs a} depicts that, for all values of $q$, the value of $T_{HP}$ is directly proportional to $a$. Note that the slope of $T_{HP}$ versus $a$ is constant for $q=0$, whereas this is not the case for $q\neq 0$. Also, Fig.~\ref{Crit Tvsq} depicts the monotonically decreasing nature of $T_{HP}$ with the increase in $q$ for all values of $a$. Thus, we can conclude that, for hairy black holes, the Hawking/Page phase transition is more pronounced when the value of $q$ is small and the value of $a$ is relatively large. Therefore, for each $q$ ($a$), a minimum (maximum) value of $a$ ($q$) exists, denoted by $a_{c}$ ($q_{c}$), below (above) which phase transition is forbidden.

The derived hairy black holes must be locally stable as well. This local stability is indicative of how a system at equilibrium reacts to minor fluctuations in thermodynamic variables. In canonical ensemble, the condition for the stability of a black hole branch is the positivity of the specific heat at a constant charge $C_{q}=T(\partial S_{BH}/\partial T)|_{q}$. It is clear from Fig.~\ref{tvszhcase2-q} that the slope of $S_{BH}-T$ plane is always positive in the thermodynamically preferred hairy black hole phase \textcircled{A}. As a result, the positive value of $C_{q}$ in phase $\textcircled{A}$ signifies the local stability of the large hairy black holes. Conversely, for the thermodynamically less favoured hairy black hole phase \textcircled{B}, $C_{q}$ is negative.

\section{Conclusions}

This work showcased an alternate class of analytical solutions for $(2+1)$-dimensional charged black holes with a scalar hair in Einstein-Born-Infeld-Scalar theory where the coupling function $f(z)$ and the form factor $A(z)$ determine the nature of solutions. In this study, we primarily focus on an interesting coupling of type $f(z)$ = $e^{-A(z)} \sqrt{1+\alpha ^2 q^4 z^4}$ with two form factors: $A(z)=-\log(1+a^2 z^2)$ and $A(z)=-a^2 z^2$. The introduction of the scalar field is observed to enrich the thermodynamics of this system. The parameter $a$ controlled the intensity of the scalar field, and both the solutions smoothly reduced to the standard BTZ-like solution with Born-Infeld-like gauge field in the limit $a \rightarrow0$. Similarly, another parameter, $\alpha$, associated with the Born-Infeld gauge field, seemed to have minimal effects on the black hole geometry and thermodynamics. The geometric analysis of both solutions presents us with the following key points:  
\begin{itemize}
    \item The curvature scalars like the Kretschmann scalar and Ricci scalar are found to be always finite and regular outside the horizon, and no additional singularity is observed due to the introduction of the scalar hair.
    \item The scalar field is real and regular outside the horizon and vanishes at the asymptotic AdS boundary. 
   \item The scalar field potential takes a usual constant value equal to the cosmological constant with a negative sign at the asymptotic boundary. It is also bounded from above by its UV boundary value between the horizon and the spacetime boundary. This satisfies the Gubser criterion for a well-defined boundary theory \cite{Gubser:2000nd}.

\end{itemize}
We then analysed the thermodynamic properties of the obtained black holes in a fixed charge ensemble, and we found some interesting results. For the first metric, only a single black hole phase appeared, which remained thermodynamically stable and favoured at all temperatures. For the second and more intriguing case of $A(z)=-a^2 z^2$, with coupling same as in the first case, we observed Hawking/Page phase transition between the large stable black hole solution and the thermal-AdS phase. We found that, for a particular value of $q$, above a critical value of the hairy parameter $a_c$, the black hole can undergo Hawking/Page phase transition. However, below $a_c$, no such transitions occurred. Similarly, for a fixed value of the hairy parameter $a$, a critical value of the charge parameter $q_c$ exists above which no Hawking/Page phase transition occurred. Subsequently, we analysed the transition temperature $T_{HP}$ as a function of $a$ and $q$ and found that $T_{HP}$ increases monotonically with $a$ and decreases with $q$. It is also confirmed that the thermodynamically favoured hairy black hole phase always has positive specific heat.

There are numerous potential avenues for expanding upon this research. We believe that if the rotation parameter can be included, it will be interesting to analyse the thermodynamic structure of the rotating charged hairy black hole. Furthermore, studying the behaviour of hairy black holes under various perturbations can give deeper insights into these hairy black holes' dynamical stability. Preliminary investigations suggest that these hairy black holes exhibit dynamic stability when subjected to scalar field perturbations. We are currently conducting further research in these areas.

\section*{Acknowledgements}
S.S. and A.D. would like to express their deepest appreciation to S. Mahapatra for his invaluable guidance which shaped this research. We are also grateful to S. S. Jena and B. Shukla for their valuable help and support.

\bibliographystyle{unsrt}
\bibliography{BI}
\end{document}